\documentclass{article}
\usepackage{graphicx}
\usepackage{authblk}
\usepackage{amsmath}
\usepackage{cite}
\usepackage{xcolor}
\usepackage[a4paper, total={6.5in, 10in}]{geometry}

\title{Status of the $W$ boson mass \\ and the future of the electroweak fit in the next decades}

\author[1,2]{Giuseppe Bozzi}
\author[3]{Matthias Schott}
\affil[1]{\small Dipartimento di Fisica, Università di Cagliari, Cittadella Universitaria, I-09042, Monserrato (CA), Italy}
\affil[2]{\small INFN Sezione di Cagliari, Cittadella Universitaria, I-09042, Monserrato (CA), Italy}
\affil[3]{\small Physikalisches Institut, Universität Bonn, Bonn, Germany}

\begin{document}


\maketitle

\begin{abstract}
A precise determination of the $W$ boson mass is an essential test for the Standard Model of particle physics: the comparison of experimental value and theoretical prediction allows to probe the internal consistency of the electroweak sector and could possibly highlight signals of New Physics. We provide a concise and up-to-date summary of past and recent measurements at lepton and hadron colliders, a discussion of the known perturbative and non-perturbative theoretical ingredients used to provide predictions for the relevant observables, and an overview of future prospects to reduce systematic uncertainties and to compare different measurements in a consistent way. We conclude with a brief discussion on the relevance of the global electroweak fit at present and future colliders.
\end{abstract}

\section{The $W$ boson in the electroweak sector}

In addition to the Higgs and fermion masses and the CKM matrix, the electroweak sector of the Standard Model is uniquely determined by fixing 3 parameters in the Lagrangian: two gauge couplings $g,g^{\prime}$ and the vacuum expectation value $v$ of the Higgs field. Electroweak observables are then expressed in terms of them, i.e. $m_W=v|g|/2,\,\, m_Z=v\sqrt{g^2+g^{\prime 2}}/2,\,\, \theta_{W}=\tan^{-1}(g^{\prime}/g)$, and the predictions can be compared to the measured values to test the validity of the Standard Model.

A particularly convenient set of experimental measurements used to fix the values of ($g,g^{\prime},v$) is given by ($G_F, \alpha, m_Z$): the Fermi constant $G_F$ is obtained from the muon lifetime~\cite{MuLan:2010shf}, the fine structure constant $\alpha$ is extracted from the anomalous magnetic moment of the electron~\cite{Hanneke:2008tm}, and the $Z$ boson mass has been measured at LEP~\cite{ALEPH:2005ab} from the $Z$ line-shape. Thanks to the high experimental precision attained for each measurement, this set minimises the parametric uncertainty in the predictions for electroweak observables.  

The well-known $m_{W}-m_{Z}$ interdependence,
\begin{equation}\label{eq:central_relation}
m^{2}_{W}\left(1-\frac{m^{2}_{W}}{m^{2}_{Z}}\right)=\frac{\pi\alpha}{{\sqrt 2}G_{F}}(1+\Delta r)\,,
\end{equation}
is established through the matching of the calculation of the muon decay width within the Fermi model~\cite{behrends1956radiative,berman1958radiative,kinoshita1959radiative}, supplemented by two-loops QED corrections to the point-like interactions\cite{van1999complete,van2000precise,seidensticker1999second}, and the full Standard Model. The $\Delta r=\Delta r(\alpha, G_{F}, m_{W}, m_{Z}, m_{H}; m_{f}; CKM)$ term incorporates radiative corrections to the tree-level result. The particular relevance of Eq.(\ref{eq:central_relation}) for precision tests of the Standard Model led to a huge effort towards accurate predictions of $\Delta r$ in the past decades.

The 1-loop contributions, computed in pioneering works~\cite{Sirlin:1980nh,Marciano:1980pb,Sirlin:1981yz}, originate from the $W$ self-energy, vertex and box diagrams, together with the related counterterms. The first contributions beyond the 1-loop result were the resummation~\cite{marciano1979weak,Sirlin:1983ys} of the ${\cal O}(\alpha\ln\frac{m_f}{m_Z})$ contributions, where $m_f$ stands for a generic fermion mass, and the resummation of effects due to fermion doublets with a large mass splitting~\cite{consoli1989effect}. QCD corrections proportional to powers of the top mass were computed at the two-loop ${\cal O}(\alpha\alpha_s)$~\cite{Djouadi:1987gn} and three-loop ${\cal O}(\alpha\alpha^2_s)$ order~\cite{Avdeev:1994db,Chetyrkin:1995ix,Chetyrkin:1995js, chetyrkin1996three}. Two-loop electroweak corrections were first obtained at ${\cal O}(\alpha^2\frac{m^4_t}{m^4_W})$ in the $m_t\gg m_H$ limit~\cite{vanderBij:1986hy}, then for arbitrary Higgs mass but in the limit of vanishing coupling constants~\cite{Barbieri:1992nz,Barbieri:1992dq,Fleischer:1993ub} and successively refined with the inclusion of subleading ${\cal O}(\alpha^2\frac{m^2_t}{m^2_W})$ contributions ~\cite{Degrassi:1996mg,Degrassi:1996ps}. The full two-loop accuracy was achieved through the calculation of the ${\cal O}(\alpha\alpha_s)$ contributions of quark pairs to gauge boson self-energies~\cite{Djouadi:1987di,Kniehl:1989yc,Djouadi:1993ss,Halzen:1990je}, two-loop diagrams with at least one closed fermion loop~\cite{Freitas:2000gg,Freitas:2002ja,Awramik:2003ee} and gauge/Higgs boson loops~\cite{Awramik:2002wn,Onishchenko:2002ve,Awramik:2002vu}. Three-loop and four-loop results at ${\cal O}(\alpha^3\frac{m^6_t}{m^6_W})$ and ${\cal O}(\alpha^2\alpha_s \frac{m^4_t}{m^4_W})$~\cite{vanderBij:2000cg,Faisst:2003px} and at ${\cal O}(\alpha\alpha^{3}_{s}\frac{m^2_t}{m^2_W})$~\cite{schroder2005four,Chetyrkin:2006bj,Boughezal:2006xk} in the large-$m_t$ limit are also available. 

The prediction of the Standard Model for the $W$ boson mass in the on-shell renormalisation scheme can be obtained from Eq.(\ref{eq:central_relation}) through an iterative procedure (given the dependence of $\Delta r$ on $m_{W}$) and reads~\cite{Sirlin:1980nh,Marciano:1980pb,Sirlin:1981yz},
\begin{equation}\label{eq:SMprediction}
m^2_W=\frac{m^2_Z}{2}\left(1+\sqrt{1-\frac{4\pi\alpha}{G_F\sqrt{2} m^2_Z}(1+\Delta r)}\right).
\end{equation}
A simple parametrisation of the full two-loop result and the leading higher-order contributions has been introduced in~\cite{Awramik:2003rn}:
\begin{eqnarray}\label{eq:mWformula}
m_{W}&=&m^{0}_{W}-c_{1}dH-c_{2}dH^{2}+c_{3}dH^{4}+c_{4}(dh-1)-c_{5}d\alpha\nonumber\\
&+&c_{6}dt-c_{7}dt^{2}-c_{8}dHdt+c_{9}dhdt-c_{10}d\alpha_{s}+c_{11}dZ\,,
\end{eqnarray}
where 
\begin{eqnarray}\label{eq:mWterms}
&dH&=\ln\frac{m_{H}}{100\,{\mathrm{GeV}}}\,\,\,\,\,\,\,\,dh=\left(\frac{m_{H}}{100\,{\mathrm{GeV}}}\right)^{2}\nonumber\\
&dt&=\left(\frac{m_{t}}{174.3\,{\mathrm{GeV}}}\right)^{2}-1\,\,\,\,\,\,\,\,dZ=\frac{m_{Z}}{174.3\,{\mathrm{GeV}}}-1\nonumber\\
&d\alpha&=\frac{\Delta\alpha}{0.05907}-1\,\,\,\,\,\,\,\,d\alpha_{s}=\frac{\alpha_{s}(m_{Z})}{0.119}-1\,.\nonumber
\end{eqnarray}

The formula singles out the dependence of the perturbative corrections on the input parameters ($m_{H},m_{t},$ $m_{Z},\alpha_{s},\Delta\alpha$), where $\Delta\alpha$ represents the leptonic and hadronic contributions to the shift in the fine structure constant (numerical values for each $c_{i}$ coefficient in Eq.(\ref{eq:mWformula}) are given in~\cite{Awramik:2003rn}). Considering a $1\sigma$ variation of each input parameter in Eq.(\ref{eq:mWformula}) around its central value, the total parametric uncertainty of $m_{W}$ is 5 MeV, with a large contribution coming from the top mass. The theoretical uncertainty from missing higher-order contributions to $\Delta r$ is estimated to be 4 MeV in the on-shell scheme~\cite{Awramik:2003rn}. A formula analogous to Eq. (\ref{eq:mWformula}) has been obtained in the $\overline{MS}$ renormalisation scheme~\cite{Degrassi:2014sxa}. In this case, the quoted theoretical uncertainty (computed by means of renormalisation scale variation) is 3 MeV, comparable to the splitting between the central values obtained with the two renormalisation schemes.

Equation \ref{eq:mWformula} can be used to determine the SM expectation of the $W$ mass by employing a global electroweak fit, where all relations of observables within the electroweak sector and the corresponding experimental values of precision observables are used. The global electroweak fit can be either used to test the overall consistency of the electroweak theory, thus probing potential new physics contributions in loops, which are not taken into account in the theoretical calculations, or to leave one (or more) parameters free, i.e. do not constrain them by their measured value, and thus find their best matching value to all other observables. We refer for a detailed discussion of global electroweak fits to the latest publications of the \textsc{HEPfit}~\cite{Ciuchini:2016sjh,deBlas:2016nqo,deBlas:2017wmn,PhysRevD.106.033003,deBlas:2022hdk} and \textsc{Gfitter}~\cite{Erler:2019hds,Flacher:2008zq, Haller:2018nnx} collaborations. 

Using the most advanced theoretical informations and the  measurements~\cite{ParticleDataGroup:2022pth} of the top-quark mass, $m_{top}=172.58\pm0.45$ GeV, the Higgs Boson mass, $m_{H}=125.21\pm0.12$ GeV, the weak mixing angle $\sin^2(\theta^{lept.}_W)= 0.2324\pm0.0012$ and the $Z$ boson mass and width, $m_Z=91.1875\pm0.0021$ GeV and $\Gamma_Z=2.4955\pm0.0023$ GeV, the \textsc{HEPfit} collaboration quotes a SM expectation of $m_W^{SM}=80.354\pm0.006$ GeV~\cite{PhysRevD.106.033003}.

\begin{figure*}[t]
\centering
\resizebox{0.7\textwidth}{!}{\includegraphics{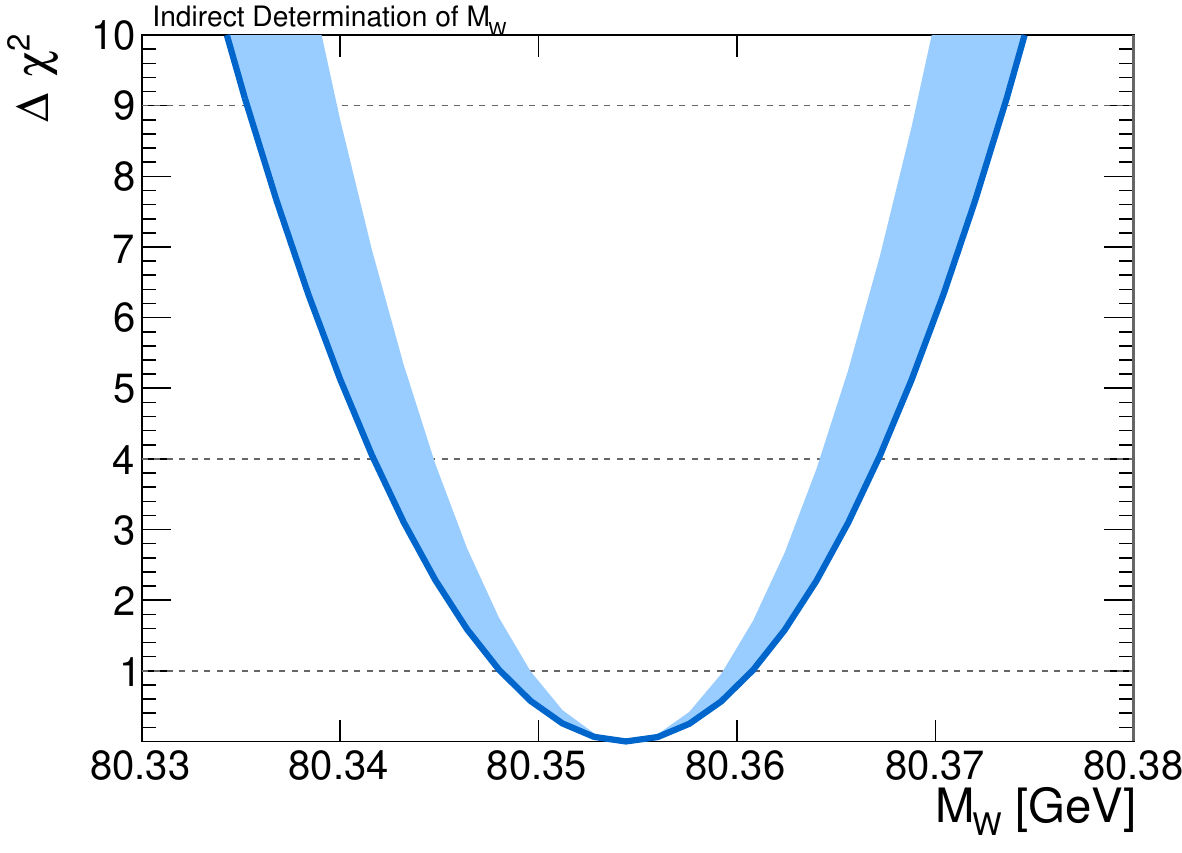}}
\caption{$\chi^2$ distribution of the global electroweak fit using the Gfitter program \cite{Flacher:2008zq} for varying values of $m_W$. Theoretical uncertainties are indicated by the filled blue areas.}
\label{fig:WMassEWFit}
\end{figure*}

The corresponding estimate by the \textsc{GFitter} collaboration, with slightly different input experimental values, is $m_W^{SM}=80.356\pm0.006$ GeV~\cite{Haller:2018nnx}. Figure \ref{fig:WMassEWFit} shows the $\Delta \chi^2$ distribution of the global electroweak fit for different values of $m_W$ including (and excluding) theoretical uncertainties, where the $\Delta \chi^2$ is defined with respect to the minimal $\chi^2$-value of the global electroweak fit, when all experimental observables are considered. 

\section{First precision measurements at the Large Electron Positron collider}

After the discovery of the $W$ boson at the UA1 and UA2 experiments~\cite{Arnison:1983rp,Banner:1983jy}, its mass was determined with a precision of approximately 5~GeV through the analysis of its decay product kinematics. 

The first accurate measurement of $m_W$ took place using electron-positron collisions during the LEP 2 runs by the four LEP experiments Aleph, Delphi, L3 and Opal~\cite{Schael:2013ita}. Electron-positron colliders have the huge advantage that the collision energy, $\sqrt{s}$, is precisely known and hence two different measurement approaches can be employed. First, the strong dependence of the production cross section of $W^+W^-$ pairs in the process $e^+e^-\rightarrow W^+W^-$ on $m_W$ close to the production threshold $\sqrt{s}\approx 2 m_W$ can be studied to determine $m_W$. Secondly, the kinematics of the decay products can be reconstructed at higher center of mass energies, where the dependence of the cross-section is weaker, but the constraint that all reconstructed energies in the detector must match the center of mass energy can still be used. This allows for the reconstruction of $m_W$ in the full hadronic decay channel (four quarks), semi-hadronic decay channel (two quarks, one charged lepton, one neutrino), and even partially the full leptonic decay channel with two charged leptons and two neutrinos in the final state. 

The LEP combined value for the $W$ boson mass ($m_W$) was determined as $m_W^{\rm LEP} = 80376 \pm 33~{\rm MeV}$. The primary systematic contribution stemmed from uncertainties in the modelling of fragmentation and hadronization processes~\cite{Schael:2013ita}. 

\section{Measurements of $m_W$ at hadron colliders}

Until 2007, LEP measurements predominantly influenced the world average, but the landscape shifted with the publication of the first precision measurement of $m_W$ at the Tevatron collider by the CDF collaboration, achieving a comparable level of precision. This prompts a detailed exploration of the principles and challenges associated with the $W$ boson mass measurement at hadron colliders, specifically delving into the measurements conducted by the CDF and D0 collaborations at the Tevatron, as well as those by the ATLAS, LHCb and CMS collaborations at the LHC.

In contrast to the LEP experiments, complete reconstruction of the $W$ boson decay kinematics in hadron collisions is limited to the transverse plane with respect to the beam axis. This limitation arises from the unknown initial collision energy of the interacting partons along the beam direction. Consequently, only the conservation of momentum in the transverse plane can be employed. 

The measurement of $m_W$ is always performed in the electron and muon decay channels, i.e. $W\rightarrow e\nu$ and $W\rightarrow \mu\nu$, as the hadronic decay channels face challenges from the overwhelming multi-jet background. The basic idea of all measurements is to relate an energy measurement of the decay products to the mass of the W boson. The three key observables that are sensitive to $m_W$ are the transverse momentum spectrum of the charged decay lepton, ${\vec p\, }_T^l$, the distribution of missing transverse energy, $\vec E_T^{\rm \,miss} = - {\vec p\, }_T^l + \vec p_T^{\,W}$, and the transverse mass distribution,
\[m_T = \sqrt{2 |{\vec p\,}_T^l|| \vec E_T^{\rm \,miss}| [ 1 - \cos \Delta \Phi ({\vec p\,}_T^l, \vec E_T^{\rm \,miss})]},\]
where $\Delta \Phi$ represents the angle between the lepton and $\vec E_T^{\rm \,miss}$ in the transverse plane with respect to the beam axis, and $\vec p_T^{\,W}$ denotes the transverse momentum of the $W$ boson in that plane. The transverse mass $m_T$ can be interpreted as the invariant mass of the dilepton system when the $W$ boson decays entirely in the transverse plane\footnote{From now on, we will omit the vector notation to improve readability.}. Experimentally, $p_T^{\, W}$ is determined by the vectorial sum of reconstructed energy clusters in the calorimeters of the detector, referred to as the hadronic recoil $u_T = \sum E^{\rm \,calo}_T$. However, the hadronic recoil only approximates $p_T^{\,W}$ and is significantly influenced by the number of simultaneous hadron collisions during a recorded event. 

\begin{figure*}[t]
\centering
\resizebox{0.49\textwidth}{!}{\includegraphics{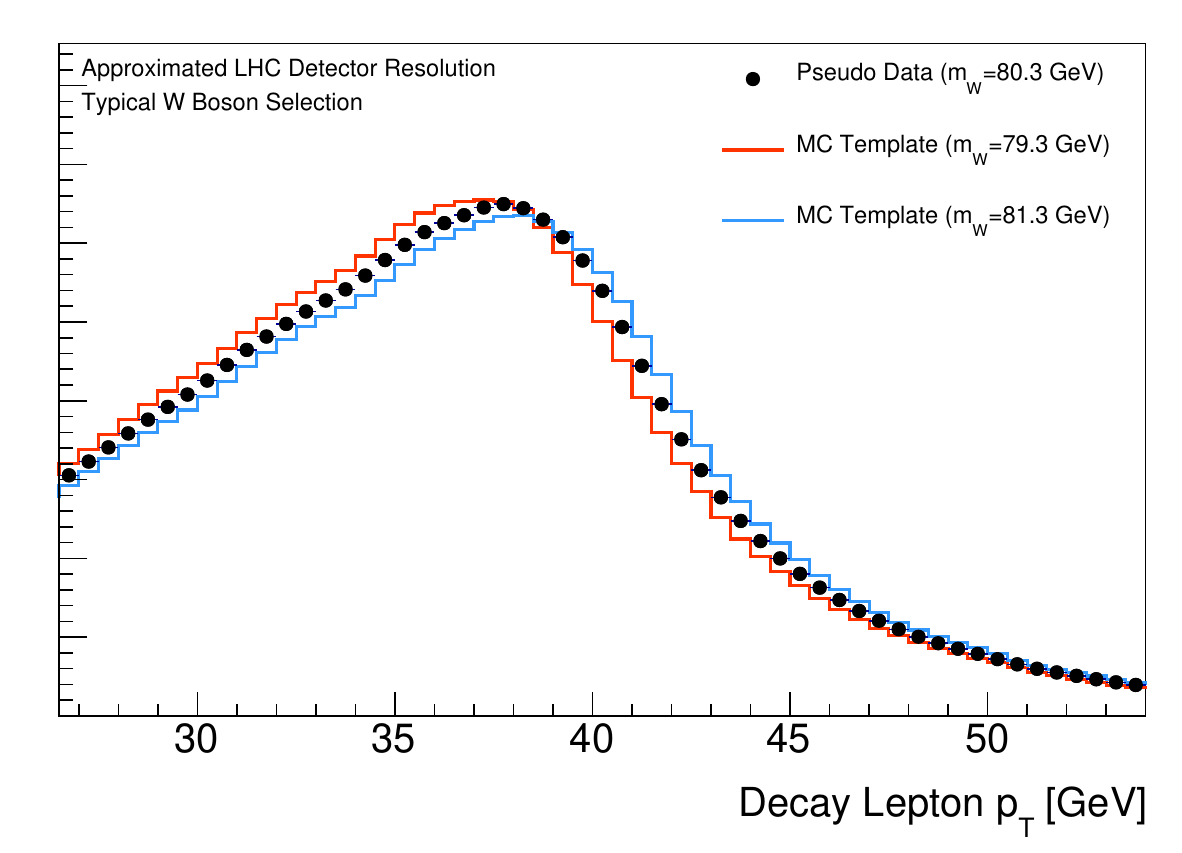}}
\resizebox{0.49\textwidth}{!}{\includegraphics{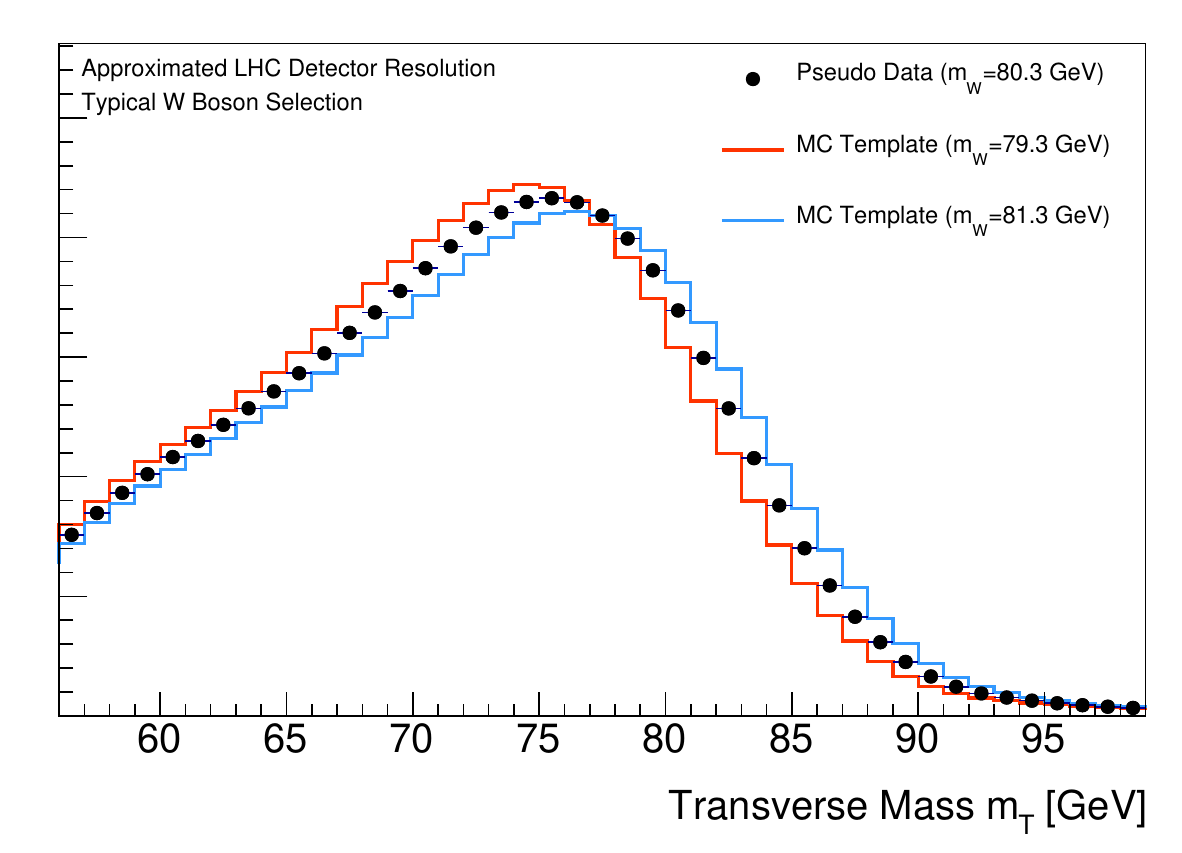}}
\caption{Templates of the $p_T^l$ (left) and $m_T$ (right) distributions for three different assumed $m_W$ masses using a typical LHC detector response, based on \textsc{Pythia8} \cite{Sjostrand:2007gs} and  \textsc{Delphes} \cite{deFavereau:2013fsa}.}
\label{fig:WMassTemplates}
\end{figure*}

Figure~\ref{fig:WMassTemplates} illustrates examples of $p_T^l$ and $m_T$ templates for three different assumed $m_W$ masses at reconstruction level of a generic LHC detector. Given the poor experimental resolution, the $E_T^{\rm miss}$ observable is typically not used for $m_W$ determinations or have only a minor influence of the final result.
These distributions rely on different aspects of the detector response and the underlying physics modelling, enabling partially uncorrelated measurements of $m_W$. Experimentally, the $p_T$ distribution is constrained by the precision of the momentum and energy scale of the tracking system and the electromagnetic calorimeter, respectively. Moreover, the modelling of the underlying transverse momentum distribution of the $W$ boson, $p_T^{\,W}$, directly impacts the transverse momentum of its decay leptons. The usage of the $m_T$ observable allows to significantly reduce the physics modelling uncertainties, but is limited by the experimental resolution of the hadronic recoil. 

Technically, the extraction of $m_W$ is realised by a template fitting procedure, where numerous theoretical predictions for $p^{l}_{T},E^{\rm \,miss}_{T},m_{T}$ are computed with different input values for $m_{W}$. For each prediction (corresponding to a definite value of $m_{W}$), the difference to experimental data is calculated, where the minimal distance defines the measured value of $m_W$. Most published results rely on the minimization of the $\chi^{2}$ value, considering statistical uncertainties only. Systematic uncertainties are included by varying the parameters determining the templates within their uncertainties, and repeating the fits. The advantage of the $\chi^{2}$-based minimization is its small computational complexity as well as the interpretability of the result. 

Alternatively, the determination of $m_W$ can also be performed through a global profile likelihood (PLH) fit \cite{Baak:2014wma}, which allows for a simultaneous optimization of $m_W$ and of nuisance parameters describing systematic uncertainties. The likelihood function, which describes the compatibility of data and MC distributions, is given by

\begin{equation}
\label{eq:plh}
L\left( \mu,\vec \theta \,| \vec n \right) = \prod_j \prod_i \text{Poisson} \left( n_{ji} | \nu_{ji} (\mu, \vec \theta) \right) \cdot \text{Gauss} \left(\vec \theta \right)
\end{equation}

where $\vec n$ represents the observed distributions in data, $n_{ji}$ is the number of events observed in bin $i$ of the distribution in a given category $j$. It is the input to the Poisson distribution with expectation $\nu_{ji}(\mu,\vec\theta) = S_{ji}(\mu,\vec\theta)+B_{ji}(\mu,\vec\theta)$, of $S_{ji}$ events from signal and $B_{ji}$ events from background contributions. The parameter of interest, $\mu$, represents variations in $m_W$ with respect to a conventional reference. Uncertainties of the signal and background distributions are represented as nuisance parameters (NPs), denoted as $\vec\theta$ in Eq.~\ref{eq:plh}, for which a normal probability distribution is assumed. The advantage of a PLH fit is that systematic uncertainties can be constrained by the same data used for the $W$ boson mass extraction, hence yielding smaller overall uncertainties. It should also be noted that the determined value of $m_W$ is expected to differ when comparing the results from a $\chi^2$ and PLH based fit approach, as the latter allows for a shift in $m_W$ to compensate for shifts of NPs.

For a precise measurement of the $W$ boson mass through a template fit approach at hadron colliders, two critical aspects must be addressed: firstly, the modelling of the detector response for the decay products of the $W$ boson must be known to high precision. Secondly, the production and decay of $W$ bosons in proton-proton collisions must be known with high accuracy, since they directly affect the kinematics of the decay leptons. Both aspects will be discussed in the following.

\subsection{Detector calibration}

Three aspects of detector effects have to be modelled in detail: the detector response on electrons and muons as well as the hadronic recoil. Lepton identification and reconstruction efficiencies are determined in data using a \textit{tag-and-probe} method in $Z$ boson events. This method relies on the complementarity of the different detector components, e.g. a muon can be reconstructed in the inner tracking detector as well as the muon tracking system, as well as the fact that the $Z$ boson decays always in same flavoured particles. For example, one requires one muon to be reconstructed in both sub-detector systems (`tag') and searches then for a track only in the inner detector system (`probe') that has opposite charge and yields together with the `tag' object an invariant mass close to the $Z$ boson. Since this implies that also the 'probe' object must be a muon, one can test if the muon tracking system also reconstructed a matching signature to the 'probe' object and therefore determine the reconstruction efficiency. Similar approaches are applied to determine also the identification, trigger and further cut-efficiencies of leptons in a multi-dimensional space, e.g. vs. the transverse momentum of the leptons $p_T^l$ or their positions in the detector. 

\begin{figure*}[t]
\centering
\resizebox{0.47\textwidth}{!}{\includegraphics{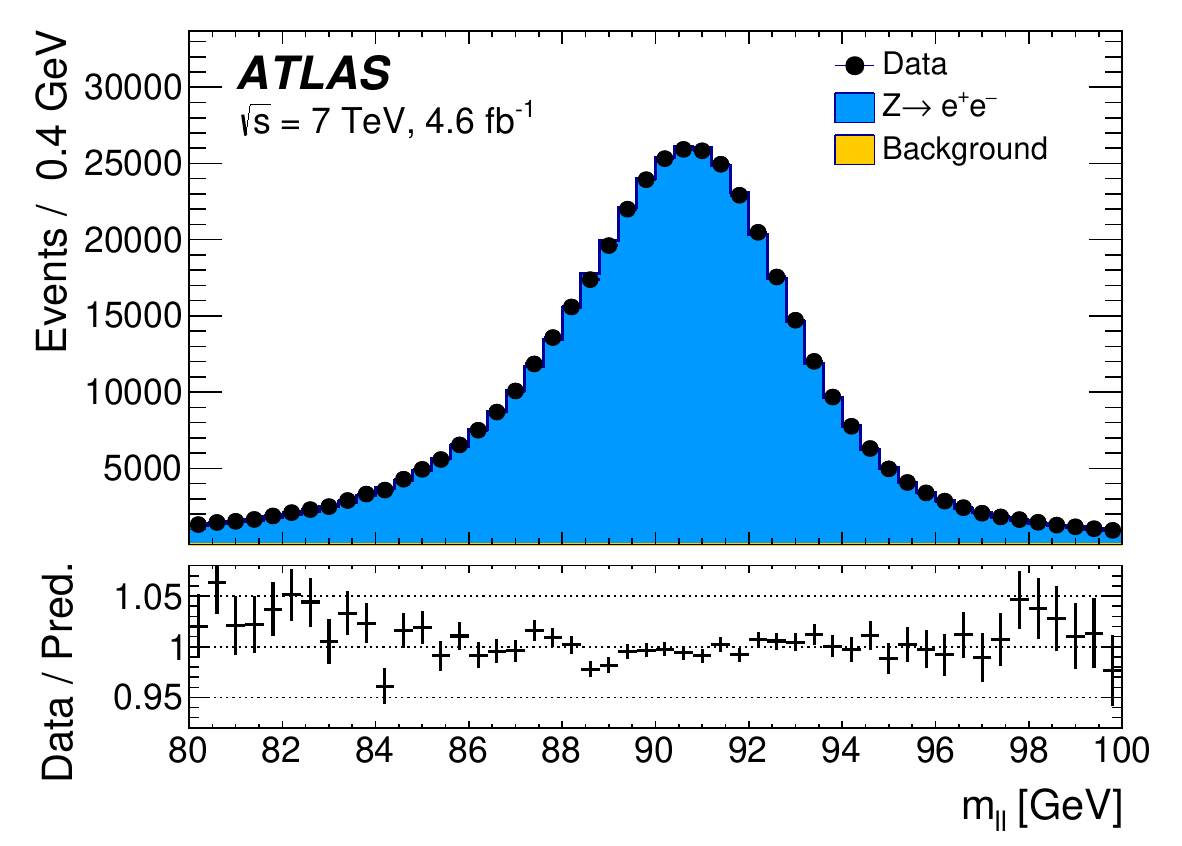}}
\resizebox{0.51\textwidth}{!}{\includegraphics{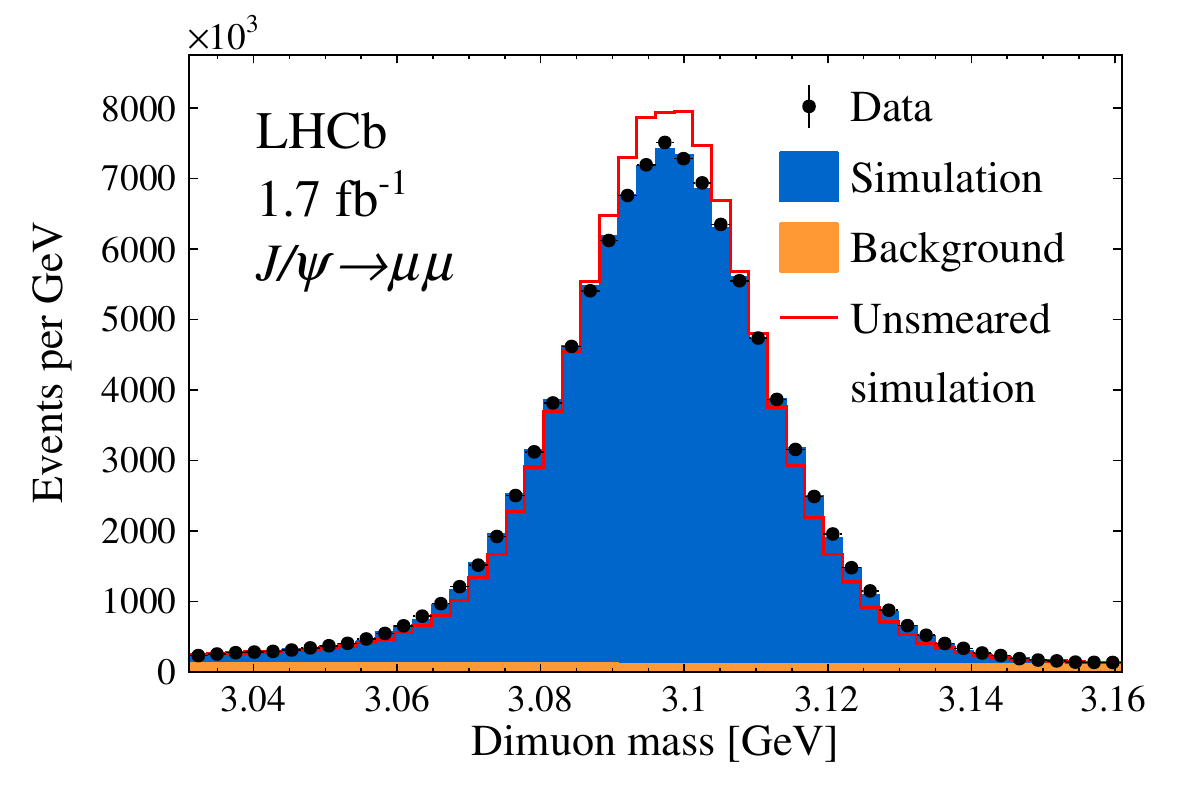}}
\caption{Dilepton mass distributions for selected $J/\Psi\rightarrow \mu\mu$ events at the LHCb Experiment (right) \cite{LHCb:2021bjt} and for selected $Z\rightarrow ee$ events for the ATLAS Experiment (left) \cite{ATLAS:2017rzl} after the lepton response calibration.}
\label{fig:DetectorCalibration}
\end{figure*}

The energy and momentum response of the detectors are typically calibrated using well known resonances, such as the $Z$ boson or the $J/\psi$, in their leptonic decay channels as their masses are known with a relative precision of 0.002\% and 0.0002\%, respectively. The invariant mass distribution of the $Z$ boson is precisely known from the LEP experiments and has the advantage that it is kinematically very close to the decay of $W$ bosons. Experimentally, one alters the detector simulated response of leptons, primarily the energy scale and resolution for $Z$ boson or $J/\psi$ events in simulated data-sets, and compares the invariant mass distribution of the dilepton systems with data. The uncorrected transverse momenta predictions in the simulation can be schematically modified via
\[
p_{T}^{MC,\,corrected} = [1+s(\eta,\phi)] \cdot p_{T}^{MC} + r(\eta, \phi) \cdot G(0,1) \cdot p_{T}^{MC},
\]
where the function $s(\eta,\phi)$ describes the corrections to the momentum scale, while the second term smears the transverse momentum randomly with a Gaussian, where the width of the Gaussian is defined by the function $r(\eta, \phi)$. The functions $s$ and $r$ are then determined in bins of relevant kinematic variables, e.g. $\eta$ and $\phi$ such that the resulting invariant mass distributions of Z boson or $J\Psi$ events match between data and corrected simulation in various categories, e.g. different energy or detector regimes. An example of the dilepton distribution for $Z\rightarrow ee$ events from the ATLAS Collaboration \cite{ATLAS:2017rzl}, as well as for $J/\psi\rightarrow \mu^+\mu^-$ from the LHCb Collaboration \cite{LHCb:2021bjt}, after the calibration procedure is shown as example in Figure \ref{fig:DetectorCalibration}.

Given that the low mass resonances, such as $J/\Psi$ or $\Upsilon$ can be precisely reconstructed with the tracking systems in their muon decay channels, but not in the electron decay channels, an alternative calibration of the calorimeter systems can be employed, making use of the ratio of the measured energy in the calorimeter and the associated electron track momenta. Here, the charged-track calibration of the inner detector system is transferred to the calorimeter using electrons from $W$ and $Z$ boson decays. This calibration procedure relies on a detailed detector simulation which includes small corrections to the amount of material upstream and downstream of the calorimeter. Then material and energy-dependent corrections are derived and applied to the simulation, followed by the determination calorimeter energy scale by a fit to the $E/p$ distribution of electrons from $W$ and $Z$ boson decays. With this approach, the CDF collaboration achieves a relative uncertainty on the energy scale of 0.007\%, which includes already the uncertainty on the charged-track momentum. 

The advantage of using the low mass resonances for the energy and momentum calibration procedures is that they allow for an independent cross-check of the $Z$ boson mass measurement. In fact, the CDF collaboration determined the $Z$ boson mass to be  $m_Z = 91192.0\pm 7.5$ MeV, which is in good agreement with the LEP legacy measurement of $m_Z = 91187.6 \pm 2.3$ MeV. Hence, the precision on the $Z$ boson mass measured at hadron colliders is already close to the one of LEP and a similar precision might be reached in the coming decade. 

Moreover, a low-mass calibration allows for a $Z$ boson independent measurement of the $W$ boson mass. This is important, as the $W$ boson as well as the $Z$ boson mass enter the same electroweak fit, where it is typically assumed that they are independent. $W$ boson mass measurements, which rely purely on a calibration based on $Z$ bosons (e.g. ATLAS and D0) are therefore actually measuring the mass ratio of both bosons. This approximation is valid in the context of the electroweak fit as long as the uncertainties of the $W$ boson mass measurements are significantly larger than the $Z$ boson mass measurement itself.

When the transverse mass of $W$ boson candidate events is used, or if selection cuts on the missing transverse energy of $W$ boson candidate events are imposed in the analysis, a maximal cut on the measured hadronic recoil in $W$ boson events is typically applied to restrict the relevant phase space to events with low transverse momenta and therefore reduce the impact of perturbative higher order corrections for multi-jet events.

Given the similarities in the production of $W$ and $Z$ bosons in hadron collisions, $Z$ bosons are also used for the detector response calibration of the hadronic recoil and therefore implicitly for the missing transverse energy. The boost of $Z$ bosons in the transverse plane of the detector has to be balanced with hadronic activity on the opposite side. Since the boost of the $Z$ boson, i.e. its transverse momentum, can be experimentally precisely determined by its decay leptons, which do not interfere with the hadronic environment, one can calibrate the reconstructed hadronic recoil such that it matches the reconstructed $p_T(Z)$ from the leptonic system. Technically, this is done by removing the reconstructed lepton signatures from all sub-detector systems, i.e. the tracking and calorimeter systems and replacing the resulting empty regions with uncorrelated detector information from other regions of the detector. The hadronic recoil is then reconstructed again using no lepton information, thus yielding similar results as a neutrino decay process. In practice, the hadronic recoil calibration is performed separately for the parallel ($u_{||}$) and transverse components ($u_{\perp}$) of the hadronic recoil vector $\vec u_T$, when projected on the $Z$ boson direction $\vec p_T^{\, Z}$.  While the $u_{||}$ distribution is directly sensitive to the energy scale, the $u_{\perp}$ is only impacted by the hadronic recoil energy resolution. The related systematic uncertainties for all detector calibrations are typically limited by the available statistics of the $Z$ boson calibration sample. 

\begin{figure*}[t]
\centering
\resizebox{0.50\textwidth}{!}{\includegraphics{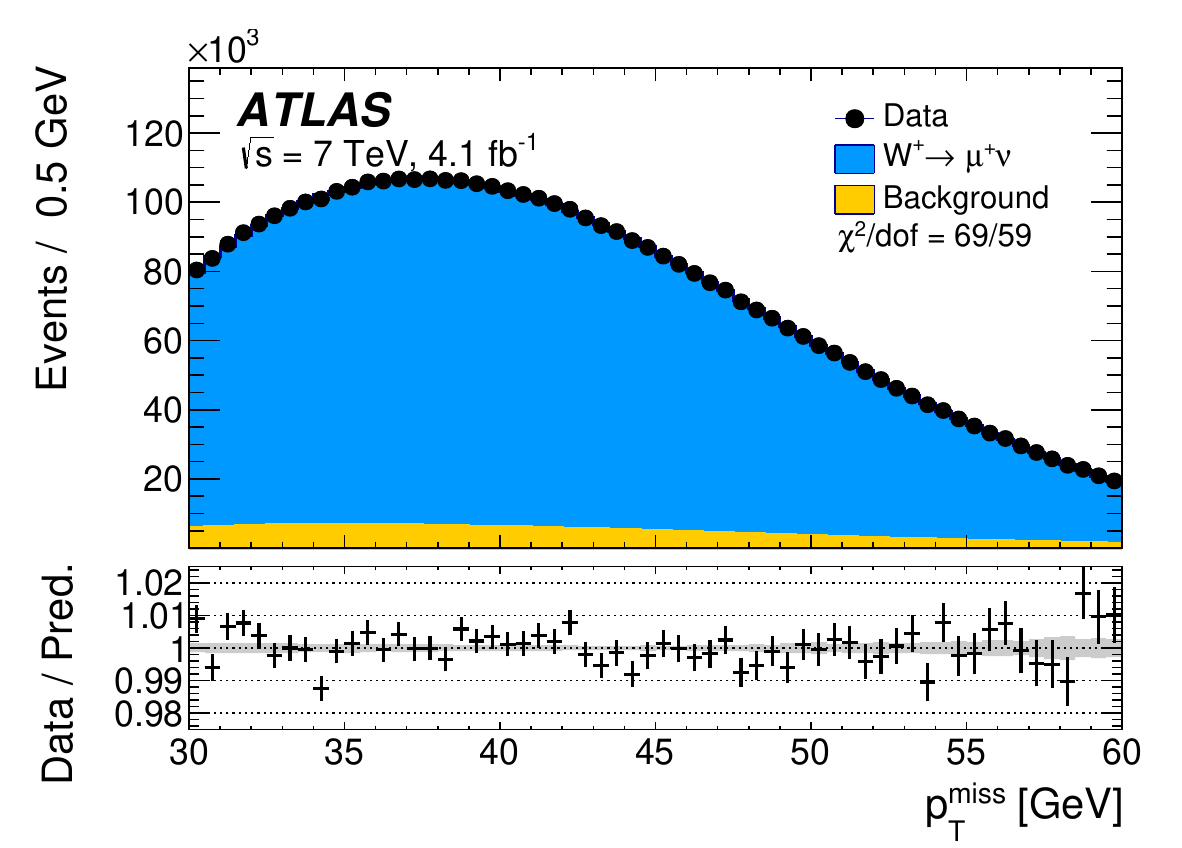}}
\resizebox{0.49\textwidth}{!}{\includegraphics{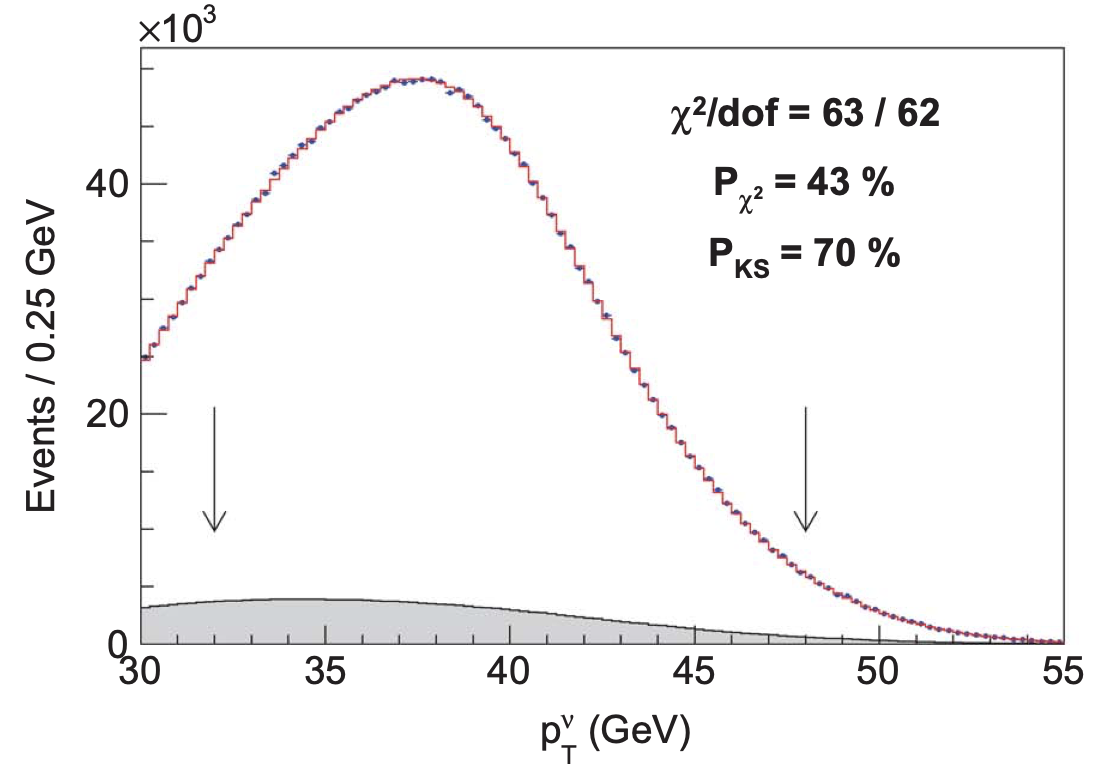}}
\caption{Missing transverse energy distributions after all corrections in the muon decay channels for the ATLAS \cite{ATLAS:2017rzl} (left) and CDF \cite{CDF:2022hxs} (right) experiments.}
\label{fig:DetectorMissingET}
\end{figure*}

The modelling of the hadronic recoil, $\vec u_T$,  directly reflects the missing transverse energy spectrum of leptonic W boson events, since they are related by $\vec u_T = \vec E^{\rm \,miss}_{T} + \vec p^{\, l}_{T}$. Examples of the measured and predicted missing transverse energy distributions after the hadronic recoil corrections are shown in Figure \ref{fig:DetectorMissingET} for the latest ATLAS and CDF analyses.

\subsection{Modelling of the W boson production}\label{modelling}
Lepton-pair production through the Drell-Yan (DY) process~\cite{Drell:1970wh,Christenson:1970um} ($q\bar q\to l^{+}l^{-}$ or $q\bar q^{\prime}\to l^{+}\nu_{l}(l^{-}\bar \nu_{l})$) is one of the most important tools for phenomenology at hadron colliders. It is a powerful test of perturbative QCD, it allows for a precise determination of the hadron structure, it provides one of the main backgrounds to New Physics searches and it is essential for precision measurements of the $W$ and $Z$ properties, including the $W$ mass.

Theoretical predictions are expressed via factorisation theorems, up to power corrections, as convolutions of collinear parton distribution functions and a partonic cross section. The target of a $m_{W}$ measurement at $10^{-4}$ accuracy ($\sim$ 10 MeV), requires extremely high accuracy for both the perturbative (EW and QCD fixed-order and all-order calculations) and non-perturbative (collinear PDFs, non-perturbative intrinsic transverse momentum of partons) ingredients.

In order to discuss the physics modelling in more detail, it is useful to decompose the fully differential DY cross section as follows,
\begin{eqnarray} 
\label{eqn:decay}
\frac{d\sigma}{d^3p_1 d^3p_2} &=& \left[ \frac{d\sigma(m_{ll})}{dm_{ll}} \right] \left[ \frac{d\sigma(y_{ll})}{dy_{ll}} \right] \left[ \frac{d\sigma(p_T,y_{ll})}{dp_T}\frac{1}{\sigma(y_{ll})} \right] \nonumber\\
&\times&\left[ (1 + \cos^2\theta) + \sum _{i=0}^7 A_i (p_T,y_{ll}) P_i(\cos\theta, \phi) \right],
\end{eqnarray}
where $p_1$ and $p_2$ are the four-momenta of the decay leptons and we consider the limit of massless leptons in a 2-body phase space and helicity conservation in the decay. 
The kinematics of the dilepton system are described by its invariant mass $m_{ll}$, its transverse momentum $p_T$ and its rapidity $y_{ll}$. 
The angles $\theta$ and $\phi$ describe the polar and the azimuthal angle of one lepton in the rest frame of the dilepton system. 
The impact of helicity and polarization effects on the decay kinematics can be described by eight spherical harmonics $P_i$ of order zero, one and two, 
weighted by eight numerical coefficients $A_i$.

The model uncertainties in these terms impact the final measurement uncertainty in $m_W$ in several ways: we will start by discussing the perturbative ingredients. 

After the pioneering~\cite{Altarelli:1979ub} work at one loop, fixed-order corrections in the QCD coupling constant have been calculated at two loops for the total cross section~\cite{Hamberg:1990np,Harlander:2002wh}, the rapidity distribution~\cite{Anastasiou:2003ds} and the fully differential cross section including leptonic decays~\cite{Melnikov:2006di,Melnikov:2006kv,Catani:2009sm,Catani:2010en}. The three loop calculation has recently been computed for the total cross section~\cite{Duhr:2020seh,Duhr:2020sdp,Duhr:2021vwj}. NLO electroweak corrections, together with mixed QCD-EW and QCD-QED corrections are also available~\cite{Baur:2001ze,Dittmaier:2001ay,Baur:2004ig,Zykunov:2006yb,CarloniCalame:2006zq,Zykunov:2005tc,CarloniCalame:2007cd,Arbuzov:2007db,Kotikov:2007vr,Kilgore:2011pa,Dittmaier:2014qza,Bonciani:2016wya,Bonciani:2019nuy,Bonciani:2020tvf,Cieri:2018sfk,deFlorian:2018wcj,Delto:2019ewv,Cieri:2020ikq,Buonocore:2021rxx,Bonciani:2021zzf,Armadillo:2022bgm,Buccioni:2022kgy}: QED final state radiation effects play the leading role in $m_W$ determination. 
At present, the highest available accuracies for fixed-order calculations are thus ${\cal O}(\alpha_{s}^{3})$ and ${\cal O}(\alpha\alpha_{s})$, while a full ${\cal O}(\alpha^{2})$ result is still missing. Subleading corrections at ${\cal O}(\alpha\alpha_s)$ are also available in SANC~\cite{Arbuzov:2005dd} and, including parton shower effects, in the WINHAC (interfaced to PYTHIA)~\cite{Placzek:2013moa} and POWHEG~\cite{Barze:2012tt,Bernaciak:2012hj,Muck:2016pko} frameworks.

Such a detailed knowledge of the mass and rapidity distribution and of the angular coefficients $A_i$ is unfortunately not enough. As already said, in charged current DY the reconstruction of the lepton-neutrino invariant mass is not possible and we need to consider observables in the transverse plane: $p^l_T, E^{\rm \,miss}_T, m_T$. The lepton transverse momentum distribution has a jacobian peak at $p^{l}_{T}\sim m_{W}/2$ (it is actually a kinematical endpoint at leading order), while the transverse mass has a kinematical endpoint at $m_{\perp}\sim m_{W}$. The sensitivity to $m_{W}$ is enhanced at these phase space edges, which are also particularly affected by the soft radiation from initial state inducing a non-zero transverse momentum of the $W$ boson. An extremely accurate prediction of the third term in Eq.~\ref{eqn:decay} is thus essential, given the ambitious goal of measuring $m_{W}$ at the 10 MeV level or less: the comparison of two theoretical templates generated with two input $m_{W}$ values differing by that amount induces a non-trivial distortion of the shapes at the permille level~\cite{Bozzi:2011ww,Bozzi:2015hha}. 

In the region of large transverse momenta, (i.e., when the transverse-momentum of the lepton pair $q_{T}$ is of the order of the invariant mass $m_{ll}$), fixed-order QCD corrections are known analytically up to $\mathcal{O}(\alpha_S^2)$~\cite{Ellis:1981hk,Arnold:1988dp,Gonsalves:1989ar,Mirkes:1992hu,Mirkes:1994dp} and numerically up to $\mathcal{O}(\alpha_S^3)$~\cite{Boughezal:2015dva,Ridder:2015dxa,Boughezal:2015ded,Boughezal:2016dtm,Gehrmann-DeRidder:2017mvr}.

However, the small-$q_T$ region is definitely the most important, since it contains the bulk of the cross section and it strongly affects the shape of the transverse observables relevant for the determination of $m_{W}$. In this region, fixed-order predictions are spoiled by initial-state soft and/or collinear radiation which causes the appearance of large logarithms of the type $\ln(q_{T}^{2}/m_{ll}^{2})$. In order to obtain reliable perturbative QCD predictions, these contributions have to be resummed to all orders in perturbation theory~\cite{Dokshitzer:1978yd,Parisi:1979se,Collins:1984kg,Bozzi:2005wk,Catani:2010pd,Monni:2016ktx}. Resummed calculations at different levels of theoretical accuracy have been performed~\cite{Bozzi:2008bb,Bozzi:2010xn,Banfi:2012du,Guzzi:2013aja,Catani:2015vma,Coradeschi:2017zzw,Bizon:2018foh,Bizon:2019zgf,Alioli:2021qbf,Camarda:2021ict}, also in the framework of Soft Collinear Effective Theory~\cite{Becher:2010tm,Becher:2011xn,Echevarria:2011epo,Echevarria:2012js,Ebert:2016gcn,Becher:2019bnm,Ebert:2020dfc,Becher:2020ugp,Billis:2021ecs,Neumann:2021zkb} and transverse-momentum dependent (TMD) factorisation~\cite{Collins:2011zzd,Collins:2012uy,Collins:2014jpa,Scimemi:2017etj,Scimemi:2019cmh,Bertone:2019nxa,Bacchetta:2019sam,Bacchetta:2022awv}. The state of the art for QCD corrections to the transverse-momentum spectrum of the DY process is represented by the full N$^{3}$LL accuracy in the low-$q_{T}$ region, consistently matched to the full NNLO accuracy in the large-$q_{T}$ region. Results at approximate N$^{4}$LL accuracy are also available~\cite{Camarda:2023dqn,Neumann:2022lft}. 

A quite large impact of final state QED radiation and a non-trivial interplay of QCD and QED corrections, have also been shown for the shape of the lepton $p_{T}$, especially in the peak region~\cite{Dittmaier:2015rxo,CarloniCalame:2016ouw,Behring:2021adr}.

We now turn to the non-perturbative ingredients entering the theoretical predictions. The limited knowledge of proton PDFs induces the largest model uncertainty in $m_W$ and impact several terms in Equation~(\ref{eqn:decay}). The uncertainty in this case is due both to the choice of different sets (and the consequent spread of best values for $m_{W}$) and to the choice of different Hessian eigenvectors or Monte Carlo replicas for a given set. 

The impact on $m_{W}$ measurements has been investigated for both the transverse mass and lepton $p_{T}$~\cite{Bozzi:2011ww,Bozzi:2015hha,Quackenbush:2015yra,Hussein:2019kqx,Gao:2022wxk}. In~\cite{Bozzi:2015zja} it has been pointed out that taking into account anti-correlation effects among central and forward detectors can lead to a reduction of the error in the combination. Including bin-bin correlation with respect to PDF variation in the fit procedure has been shown to further reduce the uncertainty~\cite{Bagnaschi:2019mzi}.

An additional uncertainty at the LHC arises from heavy-quark effects. A dedicated study~\cite{Bagnaschi:2018dnh} compared theoretical predictions performed in the 4-flavour scheme (where the $b$-quark is not included in the proton but can be produced in the final state) with respect to the 5-flavour scheme (massless-$b$ in the proton). In neutral-current DY, a non-trivial distortion on $p_T^Z$ at the 1\% level was found: the impact of this effect on $p_T^l$ in the charged-current case induces a shift for the $W$ mass in the 3-5 MeV range. Even though non-negligible and important in view of a precision measurement of $M_W$, the size of these massive-$b$ contributions suggests that using a scheme with a massless charm might instead be appropriate.

Non-perturbative contributions to transverse observables due to a partonic intrinsic-$k_{T}$, have been studied in \cite{Bacchetta:2018lna,Bozzi:2019vnl} with the inclusion of a possible flavour-dependence. The effects are comparable in size to those generated by PDF variations.

\subsection{Discussion of hadron collider measurements}

In the following, we will revise the most precise measurements of $m_W$ at hadron colliders, performed by the D0 \cite{D0:2012kms} and CDF \cite{CDF:2022hxs} collaborations in proton anti-proton collisions at the Tevatron collider at a center of mass energy of $\sqrt{s}=1.96$ TeV, as well as the ATLAS \cite{ATLAS:2017rzl, ATLAS:2023fsi}, LHCb \cite{LHCb:2021bjt} and CMS \cite{CMS:2024lrd} collaborations in proton-proton collisions at center of mass energies of $\sqrt{s}=7$ TeV and $\sqrt{s}=13$ TeV.

The analysis strategy of the given experiments differs by the used decay channels (electrons, muons), the used $m_W$-sensitive observables ($p_T$, $m_T$, $E_T^{\rm \,miss}$), their calibration procedures (low-mass resonances, $Z$ boson line-shape) as well as the used test-statistics ($\chi^2$, profile-likelihood) for the final fit. Moreover, the detector designs imply different fiducial phase-space definition, i.e. kinematic selection criteria on the decay leptons for $W$ boson candidate events. Those are summarized in Table \ref{tab:WSelection}. The CDF and D0 experiments have a smaller reach in pseudo-rapidity compared to ATLAS and CMS: however, they have better pile-up and background process conditions, which allow for lower cuts on the transverse energies. The LHCb detector is optimized for precision $b$-quark physics and hence is designed as a forward detector system, thus can only reconstruct leptons with high pseudo-rapidities.

\begin{table}[thb]
\makebox[0pt]{
\small
\begin{tabular}{|l|c|c|c|c|c|}
\hline
Experiment & D0 & CDF & LHCb & ATLAS & CMS\\
\hline
\hline
Electron & $|\eta|<$1.05 & $|\eta^e|<1.0$ & & $|\eta^e|<2.4$ & \\
& $p_T^e >$25 GeV & 30<$p_T^e<$55 GeV & & $p_T^e>$ 30 GeV & \\
\hline
Muon & & $|\eta^\mu|<1.0$ & $1.7<\eta<5.0$ & $|\eta^\mu|<2.4$    & $|\eta^\mu|<2.4$ \\
& & 30$<p_T^\mu<$55 GeV & 28$<p_T^\mu<$52 GeV & $p_T^\mu>$30 GeV & 26 $< p_T^\mu < $56 GeV \\
\hline
Neutrino & $p_T^\nu>$25 GeV & 30$<p_T^\nu<$55 GeV & & $p_T^\nu>$ 30 GeV & \\
\hline
Had. Recoil & $|\vec u|<$30 GeV & $|\vec u|$<15 GeV & & $|\vec u|<$30 GeV & \\
\hline
Trans. Mass & 50$<m_T<$200 GeV & 60$<m_T<$100 GeV & & $m_T>$60 GeV &	$m_T>$40 GeV \\
\hline
\end{tabular}
}
\centering
\caption{\label{tab:WSelection}Kinematic requirements on event observables for the selection of $W$ boson events for the different experiments.}
\end{table}

The D0 experiment is based on half of the available data set of the Tevatron Run~2  and employs only the electron decay channel \cite{D0:2012kms}. Its lepton energy and momentum calibration is based on the $Z$ boson mass peak and the extraction of $m_W$ is performed on the inclusive $p_T^l$, $m_T$ and $E_T^{\rm miss}$ distributions for leptons which are reconstructed in the central part of the detector, defined by a pseudo-rapidity requirement of $|\eta|<1.05$. Currently it is not planned to analyse the remaining part of the data-set. Default samples for $W$ and $Z$ production and decay are simulated with RESBOS \cite{PhysRevD.56.5558}, an event generator including resummation of soft gluons at the NNLL accuracy, combined with PHOTOS \cite{Golonka:2005pn} for QED radiative corrections. The input PDF set is CTEQ6.6 \cite{PhysRevD.78.013004}. A comparison with the event generators WGRAD \cite{PhysRevD.59.013002} and ZGRAD \cite{PhysRevD.65.033007} is used to assess the $m_W$ uncertainty due to QED effects. The non-perturbative modelling of the boson transverse momentum relies on the BLNY parametrisation fitted on $Z$ data \cite{PhysRevD.67.073016}: the uncertainty on each NP parameter is then propagated to the $p_T^l$, $m_T$ and $E_T^{\rm miss}$ distributions. PDF uncertainties are estimated through a template fit at 68\% CL on ensembles of $W$ events generated with the input PDF error set. The final measurements of $m_W$ by D0 yields a value of $m_W=80375 \pm 23$ MeV. 

The CDF experiment uses the full data set of the Tevatron Run~2 with an integrated luminosity of 8.8\,fb$^{-1}$ and analyses the electron and muon decay channel \cite{CDF:2022hxs}. The track momentum calibration is mainly based on $J/\Psi\rightarrow\mu\mu$ events, while the energy calibration is done using the $E/p$-ratio of $Z\rightarrow ee$ and $W\rightarrow e\nu$ events. Similar to the D0 experiment, a $\chi^2$ minimization method was used for the $m_W$ determination using both decay channels and the inclusive $p_T^l$, $m_T$ and $E_T^{\rm miss}$ distributions for leptons with $|\eta|<1.0$. The simulation codes, perturbative accuracies and input PDFs are the same used for the D0 analysis. QED uncertainties are obtained from a comparison with the HORACE code \cite{HORACE} and propagated to the relevant observables. The NP parametrisation is the same used by D0, but in this case a new fit has been performed (together with a tuning of the strong coupling) on Z data. The anti-correlation between the resulting uncertainties on $\alpha_S$ and the BLNY parameters has been used to further reduce the uncertainty stemming from NP modelling. The $p_{T}^W/p_{T}^Z$ modelling uncertainty is estimated by propagating the envelope of renormalisation, factorisation and resummation scale uncertainties obtained with the DYqT code \cite{BOZZI2011207,BOZZI2009174} at NNLL perturbative accuracy. A detailed analysis is devoted to PDF uncertainties, taking into account variations induced by the use of ABMP16 \cite{ABMP}, CJ15 \cite{CJ15}, CT18 \cite{CT18}, MMHT2014 \cite{MMHT2014} and NNPDF3.1 \cite{NNPDF3.1} sets compared to the default CTEQ6.6 set. The final measured value is $m_W= 80435 \pm 9$ MeV. 

The $m_W$ measurement of LHCb is based on an integrated luminosity of $1.7$ fb$^{-1}$, recorded in 2016 and uses only the muon decay channel, which is calibrated mainly using $Z\rightarrow \mu\mu$ events \cite{LHCb:2021bjt}. Given its detector concept, the muons are reconstructed in the forward region with $1.7<|\eta|<5.0$, hence probing a different Björken-$x$ regime in the proton-PDFs compared to central collisions. The $m_W$ value is determined through a simultaneous fit of the leptonic $p_T^l$ distribution of W boson candidates and the $\phi^*$ distribution of Z boson candidate events, where the $\phi^*$ distribution can be seen as a proxy of the transverse momentum distribution of the $Z$ boson. The fit  determines not only $m_W$ but also simultaneously several other parameters, such as the fractions of $W^+$ and $W^-$ events as well as parameters describing the transverse momentum distribution of the vector bosons. Default samples for predictions are generated with POWHEG \cite{Alioli:2008gx} + PYTHIA 8 \cite{Sjostrand:2014zea}, but alternative codes are used to estimate modelling uncertainty: PYTHIA stand-alone, HERWIG \cite{Bellm:2015jjp} stand-alone, POWHEG + HERWIG and DYTurbo \cite{Camarda:2019zyx}. In addition, renormalisation and factorisation scale uncertainties on predictions for the angular coefficients are computed with DYTurbo. The impact of QED effects is estimated through the comparison of HERWIG, PYTHIA and PHOTOS predictions of the energy difference between the final-state lepton system before and after radiation. Separate $m_W$ fits based on NNPDF3.1, CT18 and MSHT20 \cite{MSHT20} are performed, each with their own uncertainty: assuming they are fully correlated (being based on almost the same dataset), the PDF uncertainty is quoted as the arithmetic average of the three fits. It is interesting to note that the selection of events with forward leptons implies also differences in the involved PDF uncertainties, thus providing a partially uncorrelated measurement of $m_W$ in terms of physics modelling. The $W$ boson mass is found to be $80354\pm32$ MeV. 

The ATLAS measurement uses the $p_T^l$ and $m_T$ distributions in the electron and muon channel for the determination of $m_W$, using data collected in 2011 at the LHC with a center of mass energy of 7 TeV and an integrated luminosity of 4.6 fb$^{-1}$. The energy and momentum calibration is performed using $Z$ boson events and leptons are considered for the analysis up to $|\eta|<2.4$. Default samples for predictions are generated with the POWHEG event generator at NLO QCD supplemented by PYTHIA 8. A reweighing technique is used to include higher-order effects. The NLO electroweak corrections, and the associated uncertainties, are simulated with WINHAC \cite{Placzek:2013moa} supplemented by PYTHIA (for the QCD and QED initial state radiation). Concerning fixed-order predictions for the angular coefficients and rapidity distribution, PDF uncertainties are estimated with the Hessian method on the CT10nnlo set \cite{Pumplin:2001ct}, also taking into account the strong anti-correlation between $W^+$ and $W^-$ \cite{ATL-PHYS-PUB-2014-015}. The same analysis is performed with the MMHT2014 and CT14 PDF sets: the corresponding uncertainties are then summed in quadrature. Additional sources of uncertainty are the propagation of $Z$-data uncertainties used to measure the angular coefficients and the propagation of the mismatch between measurements and NNLO predictions for the $A_2$ angular coefficient \cite{ATLAS:2016rnf}. As for the parton shower predictions, uncertainties are estimated through factorisation scale variation, variation of the charm quark mass, and the propagation of experimental uncertainties affecting the values of PYTHIA parameters. The first measurement of ATLAS employs a simple $\chi^2$ minimization technique \cite{ATLAS:2017rzl} yielding a value of $m_W=80370\pm19$ MeV, using separately $p_T^l$ and $m_T$ distributions for positively and negatively charged leptons in three to four different regions in $\eta$ to reduce systematic uncertainties and to test the internal consistency of the $m_W$ measurements. An updated measurement \cite{ ATLAS:2023fsi} is based on the same data-set and calibration but uses a PLH-fit and newer PDF sets, resulting in $m_W=80360\pm16$ MeV. The PLH-fit allows therefore to constrain several systematic uncertainties and reduces the overall relative uncertainty by nearly 20\%. 

While at present the CMS Collaboration's first measurement of the W boson is still under journal review, it is summarized in the following. The measurement was performed using data collected in 2016 at the LHC, corresponding to an integrated luminosity of 16.8 fb$^{-1}$ at a center-of-mass energy of $\sqrt{s} = 13$ TeV \cite{CMS:2024lrd}. Simulated samples of W and Z boson events were generated using the {\sc MiNNLO} framework \cite{Monni:2019whf, Monni:2020nks}, interfaced with \textsc{Pythia8} \cite{Sjostrand:2014zea} for parton showering and hadronization, and \textsc{Photos++} \cite{Golonka:2005pn} for modelling final-state photon radiation. The CT18Z parton distribution function (PDF) set \cite{CT18} was employed throughout. This simulation setup provides next-to-next-to-leading order (NNLO) accuracy with next-to-next-to-next-to-leading-logarithm (N3LL) resummation for the transverse momentum ($p_T$) distributions of the decay leptons (achieved through SCETlib \cite{Ebert:2020dfc,Billis:2019vxg} + DYTurbo \cite{Camarda:2019zyx}). A central feature of the CMS analysis is its momentum calibration strategy, which is anchored in $J/\psi \to \mu^+\mu^-$ decays. Z boson events are subsequently used to constrain momentum scale uncertainties and derive efficiency corrections. However, the derived momentum scale correction uncertainties of the $J/\Psi$ decays have been enlarged to cover the measured Z boson mass. In contrast to previous measurements by ATLAS and CDF, the CMS approach is based exclusively on the muon decay channel and utilizes only the muon transverse momentum ($p_T$) distribution. Similar to ATLAS, the extraction of $m_W$ is performed via a profile likelihood (PLH) fit across a finely binned two-dimensional space defined by the muon’s transverse momentum, pseudorapidity, and charge. This method tries to use a simultaneous optimization of the $m_W$ value alongside nuisance parameters associated with systematic uncertainties, thereby enhancing the overall measurement precision. A detailed summary of the CMS results, including the breakdown of statistical and systematic uncertainties, is provided in Table~\ref{tab:WMass}. While the overall modelling and experimental uncertainties are comparable to those of the ATLAS measurement, CMS benefits from a significantly larger dataset, which helps suppress statistical uncertainties. However, the exclusive focus on the muon channel and the $p_T$ observable limits the scope for cross-checks, potentially constraining the robustness of internal validation. In addition, a still open question is whether the profiling of scale uncertainties in the CMS analysis is fully justified, or might lead to an underestimation/misinterpretation of the associated theoretical systematics.

The breakdown of systematic uncertainties for all measurements is summarized in Table \ref{tab:WMass}. It should be noted that the measurements using the $p_T^l$ and $m_T$ distributions have a very different impact on the combined measurement. While the Tevatron measurements are mainly driven by the $m_T$ distribution, the measurement of ATLAS is driven by the $p_T^l$ distribution; in fact, this is the reason why the CMS collaboration concentrated their analysis only on the $p_T^l$ spectrum as observable. The significantly smaller pile-up contribution\footnote{Pile-up refers to the number of simultaneous proton-proton collisions within one recorded event, i.e., bunch-crossing.} at the Tevatron leads to a better resolution of the hadronic recoil, and therefore a higher sensitivity of $m_T$. The PDF-related uncertainties are dominant for all measurements performed at hadron colliders, but they arise from different origins, as previously discussed.

\begin{table}[h!]
\makebox[0pt]{
\small
\begin{tabular}{|l|cc|c|c|}
\hline
Experiment & \multicolumn{2}{|c|}{D0 \cite{D0:2012kms}} & \multicolumn{2}{|c|}{CDF \cite{CDF:2022hxs}}	\\
\hline
Observable & \multicolumn{1}{|c|}{$p_T^{l}$[MeV]} & \multicolumn{1}{|c|}{$m_T$ [MeV]} & \multicolumn{1}{|c|}{$p_T^{l}$[MeV]} & \multicolumn{1}{|c|}{$m_T$ [MeV]} \\
\hline
$m_W$ &	\multicolumn{1}{|c|}{80343}	& \multicolumn{1}{|c|}{80371}	& \multicolumn{1}{|c|}{80421}	& \multicolumn{1}{|c|}{80439}	\\
\hline
Stat. Unc. & 14 & 13 & 8 & 8 \\
Sys. Unc. & 20 & 18 & 6 & 5 \\
Model Unc. & 14 & 13 & 6 & 5 \\
\hline
Total Unc. & 28 & 26 & 12 & 10 \\
\hline
Lepton Calib. Unc. & 18 & 17 & 2 & 2 \\
Had. Calib. Unc. & 6	& 5 & 4 & 2 \\
Other Exp. Unc. & 2	& 1 & 5	& 3	\\
PDF	& 11 & 11	& 4	& 4	\\
QED Effects	& 7	& 7 & 3 & 3 \\
$p_T(W)$ modelling & 2 & 5 & 2 & 1 \\
\hline
Final Result of Collaboration & \multicolumn{2}{|c|}{$80375\pm23$ \cite{D0:2009yxq}} & \multicolumn{2}{|c|}{$80433\pm9$} \\
(Stat., Exp. Sys., Model Unc.) & \multicolumn{2}{|c|}{$(\pm11\pm15\pm13$)} & \multicolumn{2}{|c|}{$(\pm6\pm5\pm5$)} \\
\hline
\hline
Experiment				&	\multicolumn{2}{|c|}{ATLAS \cite{ATLAS:2023fsi}}	&	LHCb \cite{LHCb:2021bjt} & CMS \cite{CMS:2024lrd}	\\
\hline
Observable				&	\multicolumn{1}{|c|}{$p_T^{l}$[MeV]}	&	\multicolumn{1}{|c|}{$m_T$ [MeV]}	& $p_T^{l}$[MeV] & $p_T^{l}$[MeV]	\\
\hline
$m_W$					&	\multicolumn{1}{|c|}{80360}	&	\multicolumn{1}{|c|}{80382	}&	80354	& 80360	\\
\hline
Stat. Unc.					&	5	&	7	&	23	& 2.4	\\
Sys. Unc. 					&	10	&	16	&	10	& 6.9	\\
Model Unc. 				&	11	&	20	&	19	& 6.7	\\
\hline
Total Unc. 				&	16	&	25	&	31	& 9.9	\\
\hline
Lepton Calib. Unc.			&	8	&	9	&	10	& 5.6	\\
Had. Calib. Unc.			&	1	&	12	&	-	& -	\\
Other Exp. Unc.			&	3	&	6	&	2	& 3.2	\\
PDF						&	8	&	15	&	9	& 4.4	\\
EW\&QED Effects			&	6	&	6	&	7	& 2.0	\\
$p_T(W)$ modelling			&	5	&	10	&	11	& 2.0	\\
\hline
Final Result of Collaboration	&	\multicolumn{2}{|c|}{$80360\pm16$}		&	$80354\pm31$	& $80360\pm10$		\\
(Stat., Exp. Sys., Model Unc.)	&	\multicolumn{2}{|c|}{$(\pm5\pm10\pm11)$}	&	$(\pm23\pm10\pm19)$	& $(\pm2.4\pm6.9\pm6.7)$			\\

\hline
\end{tabular}
}
\centering
\caption{\label{tab:WMass}Overview of the most precise determinations of the $W$ boson mass at hadron colliders using the lepton $p_T$ and the transverse mass $m_T$ distributions. The experimental systematic uncertainties are broken down to effects due to the lepton response calibration, the calibration of the hadronic recoil reconstruction and further experimental uncertainties such as backgrounds. Contributions to modelling uncertainties from PDFs, QED effects as well as the modelling of $p_T(W)$ are shown separately. Several uncertainties have been approximated, since detailed information within the original publications have been partly not available. The final quoted D0 number includes a combination with the previous D0 measurement.}
\end{table}

\begin{figure*}[t]
\centering
\resizebox{0.8\textwidth}{!}{\includegraphics{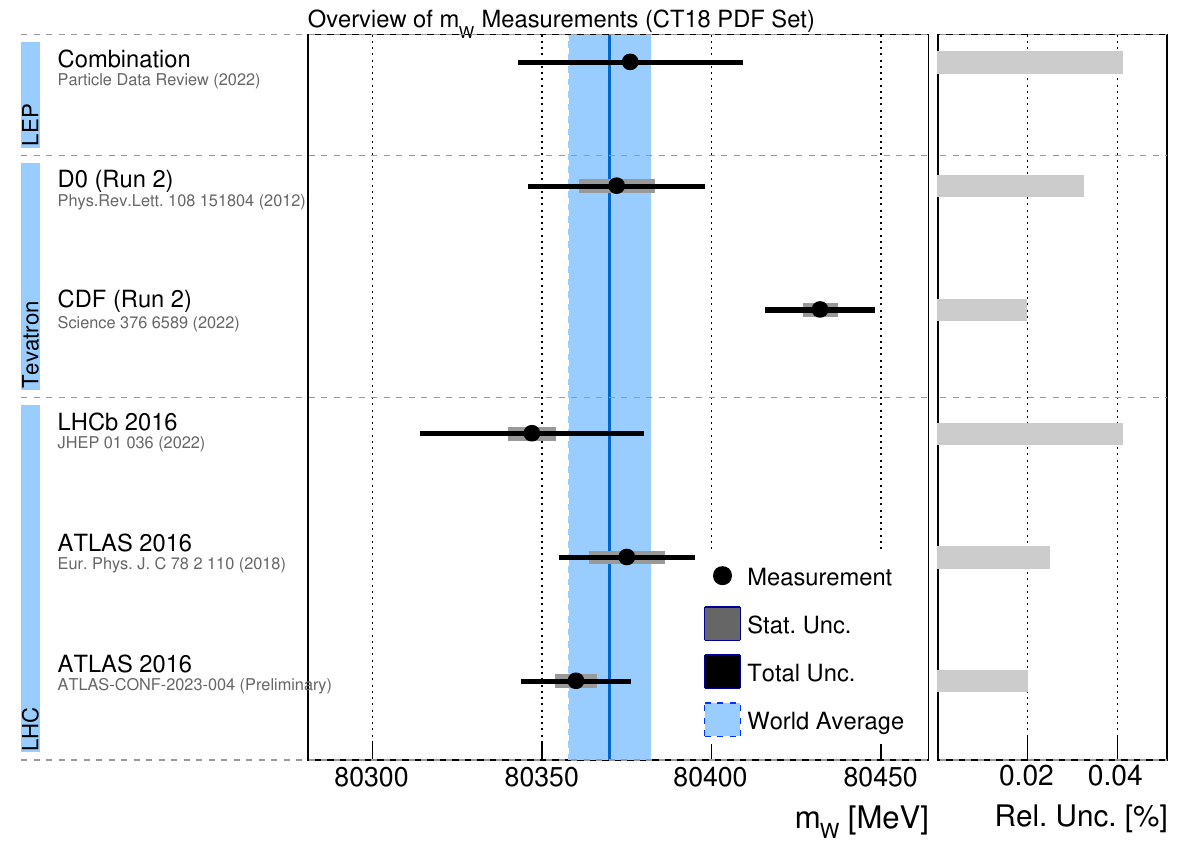}}
\caption{Summary of W Boson Measurements using the CT18 PDF set. The central values and uncertainties are based on \cite{Amoroso:2023pey} and \cite{ATLAS:2023fsi}, while the SM prediction is based on the Gfitter program \cite{Flacher:2008zq}.}
\label{fig:WMassSummary}
\end{figure*}

All measurements discussed above use different physics modelling assumptions, in particular on the angular-coefficients, the transverse momentum spectrum of vector bosons in hadron collisions and most importantly on the proton PDFs. In order to compare those measurements, they have to be transferred to the same modelling assumptions, preferably the most modern ones. This transfer was studied in detail in \cite{Amoroso:2023pey}, where also the compatibility of the different measurements has been evaluated. In particular several updates on the Tevatron measurement have been performed yielding to different central values and uncertainties. The updated ATLAS measurement \cite{ATLAS:2023fsi} was not yet included in this study. The extrapolated results are summarized in Table \ref{tab:WComb} for the CT18 PDF set, which yields to most conservative uncertainties and shown in Figure \ref{fig:WMassSummary} together with the expectation of $m_W$ in the global electroweak fit.

\begin{table}[thb]
\small
\begin{tabular}{|l|c|c|c|c|c|}
\hline
Experiment & D0 & CDF & LHCb &	ATLAS & CMS	\\
\hline
$m_W$ [MeV] & $80372 \pm 26$ &	$80432 \pm 16$ & $80347 \pm 33$ & $80360\pm16$ & $80360\pm10$ \\
\hline
\end{tabular}
\centering
\caption{\label{tab:WComb}Overview of the most precise $W$ boson mass measurements at hadron colliders. The measurements of D0, CDF and LHCb have been extrapolated to the CT18 PDF set and the same underlying modelling of the W boson production \cite{Amoroso:2023pey}. The latest measurements of ATLAS and CMS also employ a derivation of the CT18 PDF-set: however, they employ a profile-likelihood technique during the fit.}
\end{table}

It is clear that the measurement of CDF is in disagreement with the SM expectation at 4.6$\sigma$ level\footnote{Assuming the published uncertainties of the CDF measurement, the tension to the SM expectation is 7.2$\sigma$.} and with all other hadron collider measurements with a $3.6\sigma$ tension, i.e. a probability for consistency of less than $0.5\%$ \cite{Amoroso:2023pey}. However, a combination of the LHCb, D0 and the first ATLAS measurements, yields a value of 80369 $\pm$ 13 MeV using the CT18 PDF set, with a 91\% probability of consistency \cite{Amoroso:2023pey}. Combining this value with the LEP measurements results in a final experimental value of 80370 $\pm$ 12 MeV. This value is in perfect agreement with the SM expectation as well as the updated measurement of ATLAS. In this work, we also present a preliminary combination of the most recent $W$ boson mass measurements from CMS, ATLAS, LHCb, and D0, all performed using derivations of the CT18 PDF set and the \textsc{Blue} package \cite{Nisius:2020jmf}. Since the PDF uncertainties of the ATLAS and CMS measurements are profiled, the combination of the associated PDF uncertainties is not straightforward and is here only performed at an approximate level. While experimental and statistical uncertainties are treated as uncorrelated, systematic uncertainties are assumed to be highly correlated between ATLAS and CMS (correlation coefficients between 0.7 and 0.9), and moderately correlated with the other experiments (0.6 to 0.8). The resulting combined value is

\[m_W^{Average} = 80361 \pm 8\, \mathrm{MeV},\]

in excellent agreement with the Standard Model expectation of $m_W^{SM} = 80354 \pm 6$ MeV. We stress that this combination must be treated as preliminary, since a statistically sound analysis including the correct treatment of nuisance parameters in the profile likelihood fits from ATLAS and CMS is a long-term project. However, the final result is expected to be only slightly different, given that the measurements of ATLAS and CMS are dominating the average value and they are numerically very close.

While numerous papers have been published which attribute the $W$ boson mass value from the CDF experiment to new physics effects, it is then difficult to explain the disagreement to all other measurements. Given the good agreement of ATLAS, LHCb and D0, as discussed in \cite{Amoroso:2023pey}, it is also difficult to attribute the discrepancy to unknown aspects of the underlying  modelling of the signal since the same physics modelling has been studied for all experiments\footnote{This statement does not hold for the modelling of the background.}. It is interesting to note in this context that the first precision measurement from CDF \cite{CDF:2012gpf}, which uses only 20\% of the data-set, yields a value of $80400$ MeV with an approximated uncertainty of $19$ MeV after correcting for a new track momentum calibration and PDFs \cite{CDF:2022hxs}. Since the systematic uncertainties are highly correlated, the previous and the new measurement are only compatible with few percent probability. The discrepancy between the previous and the new CDF measurement is mainly driven by the muon decay channel, and might be resolved only by new measurements or an independent reanalysis of the CDF data-set. 

\section{Future prospects}

New measurements at the LHC can dramatically improve the quoted uncertainties by several means. Current measurements are driven by the $p_T^l$ distributions of the decay leptons due to the pile-up conditions in normal LHC runs, which degrade the hadronic recoil resolution. Dedicated LHC beam configuration already allowed in the past data taking in low-pileup scenarios, which have a significantly improved hadronic recoil resolution and might allow for a competitive measurement using the $m_T$ distribution. The advantage here is that several physics modelling effects impact the $p_T^l$ and $m_T$ distributions differently and thus a significant reduction is expected when combining both observables. A first measurement of the $p_T^{W}$ distribution using such low pile-up runs has been recently published \cite{ATLAS:2023llf}.

Apart from dedicated data-sets, also several analysis improvements are under study. One obvious step is to enlarge the fit complexity by including more differential distributions, such as hadronic recoil distributions or lepton rapidity distributions, thus further reducing systematic uncertainties. Another option is the definition of new $m_W$-sensitive observables: a recent idea is based on an asymmetry around the jacobian peak of the charged-lepton transverse-momentum and potentially yields a reduced sensitivity towards the $W$ boson production modelling \cite{Rottoli:2023xdc}. 

A full use of the most advanced theoretical ingredients reported in Section~\ref{modelling} would also help in further reducing modelling systematics. A possible improvement on the theory side would be the further investigation of effects related to the intrinsic-$k_T$ of partons: the availability of TMD PDF fits at high perturbative accuracy~\cite{Bertone:2019nxa,Scimemi:2019cmh,Bacchetta:2019sam,Bacchetta:2022awv} might allow to estimate the impact of these non-perturbative contributions to the extraction of $m_W$, possibly disentangling them from those stemming from collinear PDFs.

Taking all those potential future improvements into account, it seems realistic to expect overall uncertainties on $m_W$ on $6-7$ MeV level or even lower, as originally predicted already before the start of the LHC \cite{Besson:2008zs}. 

\subsection{Mandatory Consistency Tests}

The future high precision measurements at the LHC require great care in order to reduce the risk of inconsistencies of measurements across experiments; moreover, all future measurements should already prepare for a proper combination procedure during the design phase of the analysis. The latter can be realized by publishing not only the final measured value, but also the full information of the data and MC templates along with all systematic variations as well as approximated detector response functions to account for experimental effects. This would not only allow straightforward combinations, but also in-depth studies of potential tensions. Moreover, one could wonder if some internal consistency tests for future ultra-high precision measurement are required in order to be considered in a future world average. Those tests could include:
\begin{itemize}
\item separate measurements of $m_W$ for positively and negatively charged leptons in several regions of pseudo-rapidity to test the modelling of PDFs;
\item separate measurements of $m_W$ for different pile-up regimes to test the modelling of the hadronic recoil;
\item separate measurements of $m_W$ for different data-taking periods to exclude time dependent detector effects;
\item separate measurements of $m_W$ using the $p_T^l$ and $m_T$ channel to test physics and experimental effects;
\item separate measurements of $m_W$ using the electron and muon decay channel to test experimental effects;
\item measurements of $m_Z$ using $Z\rightarrow l^+l^-$ samples, where one lepton is treated as a neutrino using the $m_W$-based template techniques in order to test experimental effects and the validity of the QCD modelling.
\end{itemize}
It is clear that some of those tests would require additional work and would not yield a higher precision of the final results: for example, the fits based on $m_T$ will be subdominant in high pile-up conditions or the momentum calibration of the muon might be experimentally easier. Nevertheless, the underlying determination of $m_W$ is a highly complex endeavor where mistakes can easily be made without being noticed. Those tests could point out those mistakes and would significantly strengthen the credibility of a published result. 

\subsection{Twilight of the Global Electroweak Fit at the LHC}

The global electroweak fit has long served as a cornerstone of precision tests of the Standard Model (SM). Before the discovery of the top quark and the Higgs boson, it successfully provided estimates of their masses based on precision measurements of other electroweak observables. Since the Higgs discovery in 2012, the fit has evolved into a precision framework to search for indirect signs of new physics by comparing estimated and measured values of key parameters: the W boson mass ($m_W$), the Z boson mass ($m_Z$), the top quark mass ($m_{\text{top}}$), and the effective weak mixing angle ($\sin^2\theta_{\text{eff}}$). Most input parameters are known with exquisite precision, and are now treated as fixed in the fit, making the remaining few powerful indicators of potential deviations from the SM.

The W boson mass has always played a special role in the fit. Its estimated value has historically had smaller uncertainties than its direct experimental measurements, making it a particularly sensitive probe for tension between the theory and experiment.

The precision of the electroweak fit is limited primarily by theoretical uncertainties—particularly due to missing higher-order corrections—and the experimental uncertainties of selected input parameters. For instance, the uncertainty for $m_W$ depends sensitively on the top-quark mass, and vice versa. Similarly, uncertainties in $\sin^2\theta_{\text{eff}}$ are affected both by missing corrections and by experimental limitations.

Originally, the electroweak fit aimed to reveal significant deviations between SM expectations and measurements, offering a clear path to new physics. Today, however, all key observables agree with the SM within uncertainties on $1\sigma$-level. This leads to an important question: what is the role of future high-precision measurements if no significant tension remains?

\begin{figure*}[t]
\centering
\resizebox{0.49\textwidth}{!}{\includegraphics{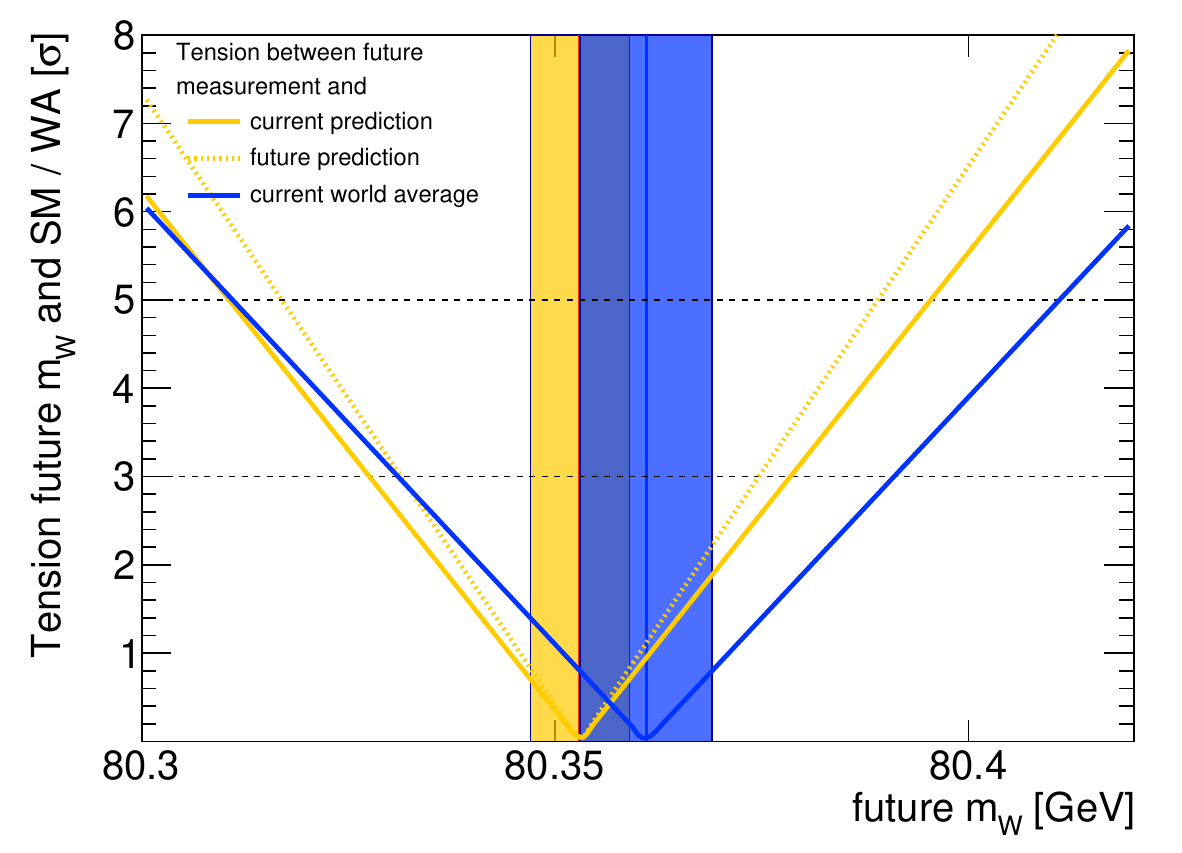}}
\resizebox{0.49\textwidth}{!}{\includegraphics{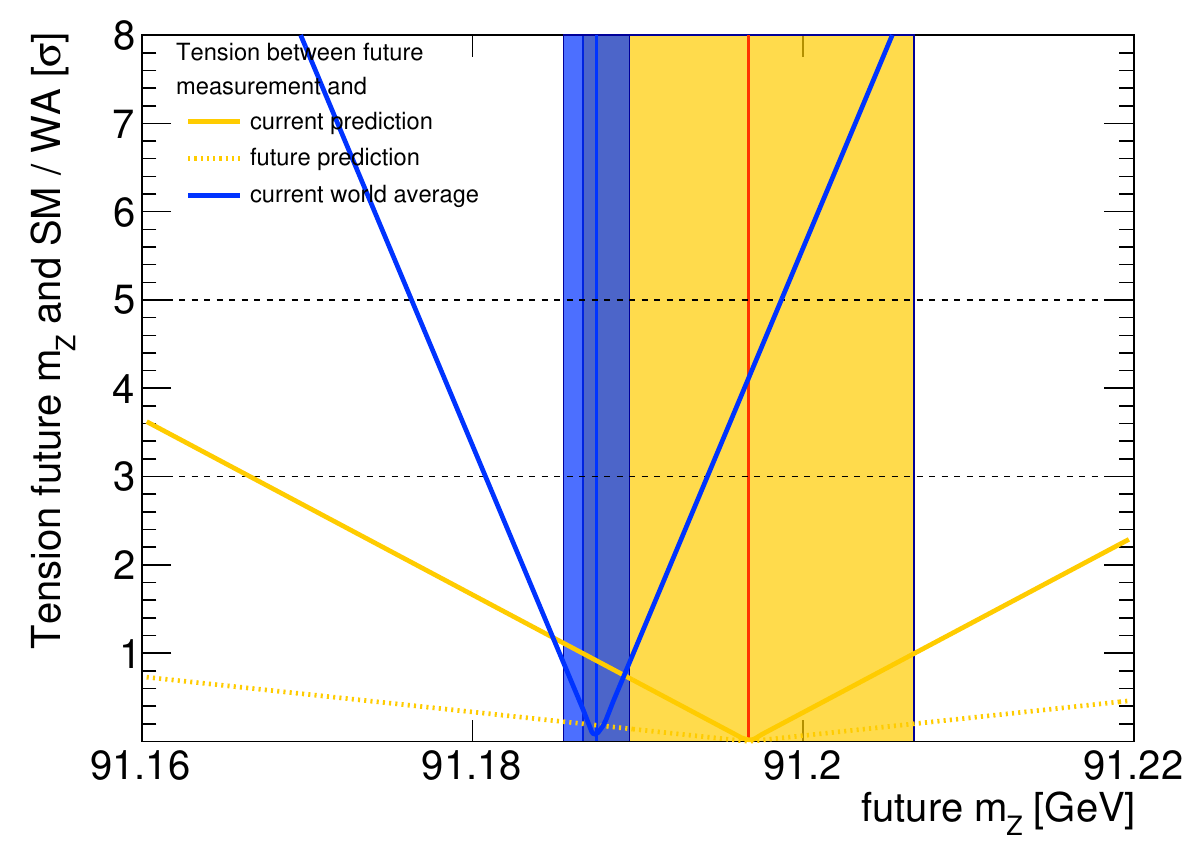}}
\resizebox{0.49\textwidth}{!}{\includegraphics{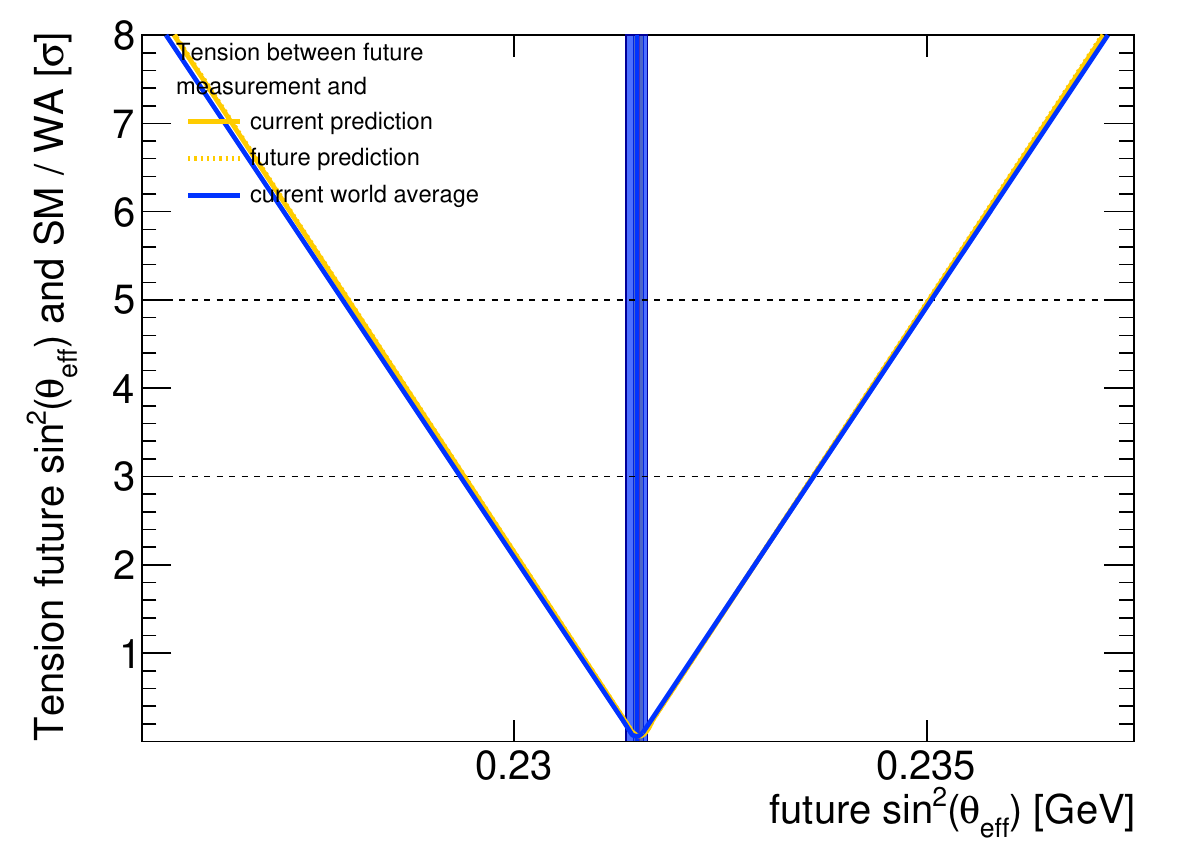}}
\resizebox{0.49\textwidth}{!}{\includegraphics{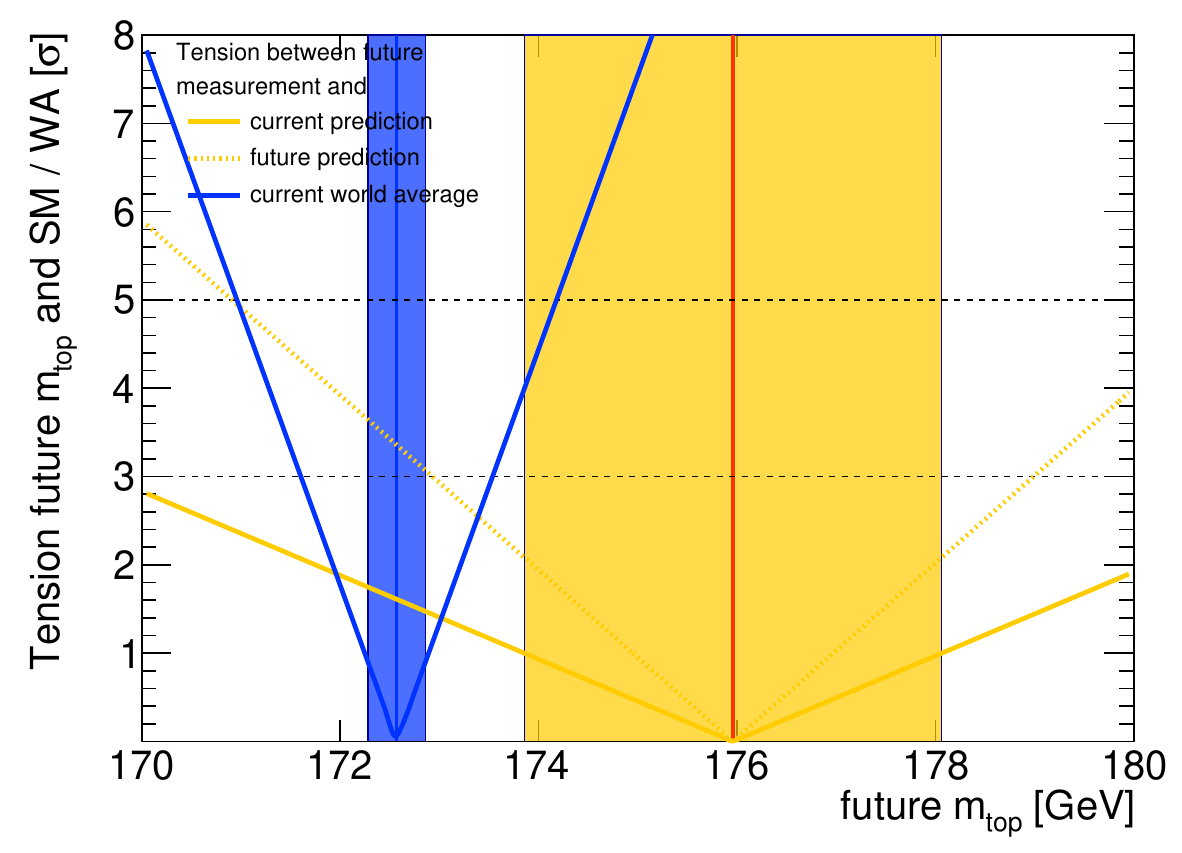}}
\caption{Dependence of the tension between the SM expectation and the current world average and a future measurement of $m_W$, $m_Z$, $\sin^2\theta_{\text{eff}}$ and $m_{\text{top}}$. The current world average (and its uncertainty) is indicated by the blue vertical line (bar), while the current expected value within the SM (and its uncertainty) is indicated by the red vertical line (yellow vertical bar). The underlying input parameters are summarized in Table \ref{tab:inputs}.}
\label{fig:EWFitDiscovery}
\end{figure*}

We argue that the potential of precision electroweak tests to reveal new physics in a model-independent way may be nearing its twilight. Any future measurement of $m_W$, $m_{\text{top}}$, or $\sin^2\theta_{\text{eff}}$ at the LHC that shows a significant deviation from the SM would also likely be in tension with the global average of existing measurements. Figure \ref{fig:EWFitDiscovery} illustrates this point by showing the significance of the tension between hypothetical future measurements and both the SM expectation and the current world average, using two scenarios: one based on current uncertainties and one assuming improved experimental precision, which also improves the uncertainties of the expected value. The corresponding input values for these projections are summarized in Table \ref{tab:inputs} and have been derived using the \textsc{Gfitter} package \cite{Haller:2022eyb}. Essentially, even when future measurements reach uncertainties which are only half as large as the current world average, there is no room for a scenario, where the new measurement would allow for a 5$\sigma$ deviation from the SM, but have less than a 2-3 $\sigma$ tension to the current world averages. Clearly, one needs to take into account that theoretical uncertainties may be underestimated in current high-precision measurements at the LHC. Hence future studies could slightly alter the current interpretation, but the overall conclusion is expected to remain largely stable.

\begin{table}[ht]
\small
\centering
\begin{tabular}{|l|c|c|c|c|}
\hline
\textbf{Observable} 	& $m_W$ [MeV] & $m_Z$ [MeV]& $\sin^2\theta_{\text{eff}}$ & $m_{\text{top}}$ [GeV]\\
\hline
& \multicolumn{4}{|c|}{\textbf{Current Precision}}\\
\hline
Measurement & 80361$\pm$8 & 91188$\pm$2 & 0.23149$\pm$0.00013 & 172.57$\pm$0.29 \\
Expectation & 80355$\pm$6 & 91192$\pm$6 & 0.23154$\pm$0.00006 & 173.82$\pm$1.50 \\
\hline
& \multicolumn{4}{|c|}{\textbf{Potential Future  Precision}}\\
\hline
Measurement & 80361$\pm$5 & 91188$\pm$1 & 0.23149$\pm$0.00007 & 172.57$\pm$0.15 \\
Expectation & 80355$\pm$5 & 91193$\pm$5 & 0.23153$\pm$0.00005 & 173.80$\pm$1.10 \\
\hline
\end{tabular}
\caption{Current and potential future precision measurements of key electroweak observables and their expected values from the global electroweak fit, with the respective observable excluded from the fit. The central values of $m_Z$, $\sin^2\theta_{\text{eff}}$ and $m_{\text{top}}$ are taken from \cite{ParticleDataGroup:2022pth}.}
\label{tab:inputs}
\end{table}

\subsection{A new era of Electroweak Precision Physics in the coming decades}

The global electroweak fit has guided the field of high-energy physics for decades, successfully predicting fundamental parameters and providing tight constraints on new physics. As previously discussed, the remarkable agreement between current measurements and SM expectations implies the waning of its power as a discovery tool in the classical sense. The era of using precision observables solely to expose internal tensions within the SM might be reaching its limits. 

Yet this does not imply that precision measurements have lost all their utility, neither in the high-luminosity era of the LHC nor beyond. In the framework of Effective Field Theory (EFT), precision electroweak observables can constrain possible extensions of the SM in a systematic, model-independent way. EFT provides a powerful tool to parametrise new physics effects through higher-dimensional operators, whose coefficients — known as Wilson coefficients — quantify the strength of new interactions at energies beyond current experimental reach. As a concrete example, the uncertainty on the W boson mass can be used to constrain Wilson coefficients in the Standard Model Effective Field Theory. At leading order, $m_W$ is sensitive to four such coefficients via the relation \cite{Dawson:2019clf}:
\[
\Delta m_W = \frac{v^2}{\Lambda ^2}(-29.827\cdot C_{\Phi l} + 14.914\cdot C_{ll} - 27.691\cdot C_{\Phi D} - 57.479\cdot C_{\Phi W B} )\, {\mathrm{GeV}}
\]

Here, $v$ is the Higgs vacuum expectation value, $\Lambda$ denotes the scale of new physics, and $C_i$ are dimensionless Wilson coefficients. While next-to-leading order corrections introduce additional operators, their numerical impact is generally subdominant. Assuming an overall precision of 10 MeV on $m_W$, one can constrain the coefficient $C_{\Phi WB}$ to below approximately $0.0025/\text{TeV}^2$. If new physics induces $O(1)$ values for these coefficients, this would imply a lower bound on the new physics scale of $\Lambda > \sqrt{1 / 0.0025} \approx 20$ TeV. Notably, the sensitivity only improves with the square root of the measurement precision: a factor-of-two gain in $m_W$ precision increases the reach in $\Lambda$ by only a factor of $\sim$1.4. This illustrates the continuing — but more nuanced — power of high-precision measurements: while they may not unearth new physics through internal inconsistencies in the SM, they remain valuable for straining physics beyond the SM via indirect, EFT-based approaches during the high luminosity era of the LHC.

The role of the global electroweak fit is poised for a fundamental transformation in the era of future electron-positron colliders. If facilities such as the FCC-ee \cite{Blondel:2021ema} or CEPC \cite{CEPCStudyGroup:2018ghi} are realized, they will enable a re-measurement of the classical electroweak precision observables — such as the $Z$ boson mass, the $W$ boson mass, and the effective weak mixing angle — with unprecedented accuracy. Dedicated runs at the $Z$ pole and the $WW$ threshold are projected to reach uncertainties at the level of $\Delta m_Z < 0.1~\mathrm{MeV}$, $\Delta m_W < 0.3~\mathrm{MeV}$ and $\Delta \sin^2\theta_{\text{eff}}^\ell < 0.00001$, improving upon LEP-era precision by more than 1-2 orders of magnitude \cite{dEnterria:2016fpc}. However, such dramatic experimental improvements must be matched by corresponding advances in theoretical predictions and uncertainty control within the global electroweak fit framework. In particular, radiative corrections, higher-order loop calculations, and parton distribution function uncertainties must be revisited and refined to avoid becoming the limiting factors in the precision program.

If theoretical uncertainties can be systematically reduced to match this new experimental precision, the global electroweak fit will once again become a sensitive probe for indirect signs of new physics beyond the Standard Model (BSM). Just as the original electroweak precision program constrained the Higgs boson mass long before its discovery, a next-generation global fit may reveal subtle tensions or deviations that point the way to new phenomena. In this sense, we would be entering a truly new era of precision physics — one in which the electroweak sector could once again serve as a leading window into uncharted territory.

\section{Summary}

The W boson mass remains a cornerstone observable in testing the electroweak sector of the Standard Model. Its precise measurement and theoretical prediction have historically provided deep insights into the consistency of the SM and the potential presence of new physics. With the new and new measurements at the LHC in the near future, a total experimental uncertainty of 8 MeV seems to be in reach, yielding a comparable precision as the estimated value of the global electroweak fit. While the classical interpretation of precision electroweak fits appears to be approaching its limits — with current measurements in remarkable agreement with SM expectations — the role of $m_W$ and related observables is far from obsolete. In the era of high-precision physics, these measurements gain renewed importance as tools to constrain effective field theory extensions of the SM.

\section*{Acknowledgements}
G.B. would like to thank A. Vicini for the fruitful collaboration and the uncountable discussions on the topic, and acknowledges support by the European Union “Next Generation EU” program through the Italian PRIN 2022 grant n.20225ZHA7W.

\bibliographystyle{unsrt}
\bibliography{References.bib}

\end{document}